\documentclass{article}
\usepackage{amsmath}        %math
\usepackage{empheq}
\usepackage{mathtools}      %work with package amsmath
\usepackage{amssymb}        %mathsymbols, say arrows, relations symbols
\usepackage{mathrsfs}       %for \mathcal, \mathscr
\usepackage{stmaryrd}       %symbols, say \boxempty
\usepackage{bm}             %bold symbols in math mode
\usepackage{amsfonts}       %blackboard bold letters, fraktur letters
\usepackage{physics}		%\ket, \bra, \ketbra, \derivative
\usepackage{cite}           %\cite
\usepackage{url}            %\url
\usepackage{hyperref}       %\autoref, \pageref, \ref

\usepackage{enumitem}       %enumerate environment,
\usepackage{amsthm}         %theorem environments
\usepackage{thmtools}       %work with package amsthm
\usepackage{nicematrix}     %for block matrix
\usepackage{geometry}
\usepackage{spalign}
\usepackage{booktabs}   %for \toprule \midrule \bottomrule
\usepackage{subcaption}
\usepackage{threeparttable}
\usepackage{multirow}

\theoremstyle{plain}
\newtheorem{theorem}{Theorem}[section]
\newtheorem{lemma}[theorem]{Lemma}
\newtheorem{corollary}[theorem]{Corollary}

\theoremstyle{definition}

\newtheorem{example}{Example}[section]

\newcommand\set[2]{ \left\{ {#1} \; \middle| \; {#2} \right\} }
\newcommand\setnd[1]{ \left\{ #1 \right\} }

\newcommand\card[1]{\left|{#1}\right|}

\newcommand{\F}{\mathbb{F}}

\newcommand{\defeq}{\triangleq}
\newcommand{\dual}[1]{#1^\bot}

\DeclareMathOperator{\wt}{\mathrm{wt}}
\DeclareMathOperator{\spn}{\mathrm{span}}
\DeclareMathOperator{\evmap}{\mathrm{ev}}
\DeclareMathOperator{\GRS}{\mathrm{GRS}}

\makeatletter  %for \autoref producing numbering with bracket
\let\oldtheequation\theequation
\renewcommand\tagform@[1]{\maketag@@@{\ignorespaces#1\unskip\@@italiccorr}}
\renewcommand\theequation{(\oldtheequation)}
\makeatother   %for \autoref producing numbering with bracket

\title{On Subcodes of the Generalized Reed-Solomon Codes}
\author{Yu Ning \\ sirning@mail.ustc.edu.cn}
\begin{document}
\maketitle
%\begin{abstract}
%In this paper, we study a class of subcodes of the generalized Reed-Solomon (GRS) codes. 
%We show equivalent characterizations for this class of subcodes of the GRS codes being self-dual, and known 
%explicit constructions of self-dual GRS codes can be exploited conveniently 
%to the construction of self-dual subcodes of GRS codes. 
%The minimum distance of the dual codes are also studied and equivalent characterizations for the subcodes of the GRS 
%codes being near-MDS are proposed. In some cases, the subcodes of the GRS codes are guaranteed to be near-MDS or MDS.
%Finally, we study the dual codes of the subcodes of the GRS codes. In some cases, the subcodes of the GRS codes can be 
%closed under taking dual codes. In other cases, the dual codes of the subcodes of the GRS codes 
%turn out to be the twisted GRS codes.
%\end{abstract}
\begin{abstract}
In this paper, we study a class of subcodes of codimension $1$ in the $[n,k+1]_q$ generalized Reed-Solomon (GRS) codes, 
whose generator matrix is derived by removing the row of degree $k-r$ from the generator matrix of the 
$[n,k+1]_q$ GRS codes, where $1 \le r \le k-1$. We show equivalent characterizations for 
this class of subcodes of the GRS codes being self-dual or near-MDS, which extends the results for $r=1$ in the 
literature. Along with these characterizations, families of self-dual near-MDS subcodes of the GRS codes are also 
proposed. Finally, for $r = 1,2$, the dual codes of the subcodes of the GRS codes are found out. 
In some cases, the subcodes of the GRS codes can be closed under taking dual codes. In other cases, the dual codes 
turn out to be the twisted GRS codes.
\end{abstract}

\section{Introduction}
Let $q$ be a prime power and $\F_q$ be the finite field with $q$ elements. 
An $[n,k,d]_q$ code is a $\F_q$-linear subspace in $\F_q^n$ of dimension $k$ and 
minimum Hamming distance $d$. For $[n,k,d]_q$ codes, the Singleton bound states that 
$d \le n-k+1$ \cite[Ch. 2]{CodeBook}, and $[n,k,n-k+1]_q$ codes reaching the Singleton bound are 
called \emph{maximum distance separable} (MDS). MDS codes are optimal in the sense that 
they have the largest minimum distance for fixed code length and dimension. 
MDS codes are also connected with objects in various research fields, such as 
orthogonal arrays \cite{OABook} in combinatorics and threshold schemes \cite{SSMDS1,SSMDS2} in cryptography. 
For an $[n,k]_q$ code $C$, the dual code of $C$ is defined as 
$$
\dual{C} = \set{\mathbf{w} \in \F_q^n}{\mathbf{w} \cdot \mathbf{c} = 0, \, \forall \mathbf{c} \in C},
$$
where $\mathbf{w} = (w_1,\dots,w_n)$, $\mathbf{c} = (c_1,\dots,c_n)$, and $\mathbf{w} \cdot \mathbf{c} = 
\sum_{i=1}^n w_ic_i$. It is known that the dual code of an MDS code is again MDS \cite[Ch. 2]{CodeBook}.
%\cite[Theorem 2.4.3]{CodeBook}. 
As a weakening of the MDS condition, $[n,k,n-k]_q$ codes are called \emph{almost-MDS (AMDS)}, and 
$C$ is called \emph{near-MDS (NMDS)} if both $C$ and $\dual{C}$ are AMDS \cite{NMDS,AMDS}. 
Since their proposition \cite{RSpaper}, the generalized Reed-Solomon (GRS) codes have become 
one of the most famous class of MDS codes with flexible parameters, and found wide applications in 
storage systems \cite{RSApplication}. Still, it is very interesting and important to study 
MDS codes not equivalent to the GRS codes or NMDS codes, and such codes may have applications in cryptography systems 
\cite{Beelen2018ISITtwistedRS,CryptanalysisTGRS} or combinatorial designs 
\cite{NMDS,DesignDing,DesignHeng}. We recall some results in the literature toward 
this direction of research:
\begin{itemize}
\item Roth and Lempel \cite{Roth1989nonGRS} considered a class of non-GRS 
codes, now called the Roth-Lempel codes, which are derived by extending the 
columns of the generator matrix of the GRS codes. 
And the characterization of the Roth-Lempel codes being MDS is related to a 
subset sum problem over finite fields 
\cite{Roth1989nonGRS}. Recently, Han and Fan \cite{RLNMDS} further 
characterized when the Roth-Lempel codes are NMDS, and 
variants of the Roth-Lempel codes were considered in \cite{RLVariant}. 
\item Another widely studied class of non-GRS codes is the 
twisted generalized Reed-Solomon (TGRS) codes introduced in 
\cite{Beelen2017ISITtwistedRS,Beelen2018ISITtwistedRS,Beelen2022ITtwistedRS}. 
Following this line of research, TGRS codes which are non-GRS MDS or 
NMDS are constructed 
\cite{Beelen2022ITtwistedRS,MDSLCDTGRS,MDSNMDSTGRS,MDSNMDS2MDSTGRS,AClass}. 
The dual codes of the TGRS codes are studied in 
\cite{parity_check,l_twists}, and self-dual TGRS codes are constructed in 
\cite{MDSNMDS2MDSTGRS,l_twists,AClass,Huang2021twistedRS}.
\item Recently, Han and Zhang \cite{Han2024NMDS} considered subcodes of codimension $1$ in the $[n,k+1]_q$ 
GRS codes, whose generator matrix is derived by removing the row of degree $k-1$ from the generator matrix of 
the $[n,k+1]_q$ GRS codes. It was shown that these subcodes of the GRS codes were either NMDS or MDS \cite{Han2024NMDS}, 
and self-dual codes were constructed out of these subcodes of the GRS codes \cite{Han2024NMDS}. 
Similarly, this kind of subcodes of the extended GRS codes were studied in \cite{subcodeEGRS}. 
More generally, Jin \emph{et, al}. \cite{Jin2024nonRSMDS} studied subcodes of the GRS codes, whose generator matrix is 
derived by removing any row from the generator matrix of the GRS codes. It was shown that these codes were non-GRS 
and equivalent conditions for them being MDS were proposed \cite{Jin2024nonRSMDS}. 
Many non-GRS MDS codes were also constructed \cite{Jin2024nonRSMDS}.
\end{itemize}
Motivated by these works, we aim to continue the study on the subcodes of the GRS codes. 
More precisely, let $\GRS_{k}(\mathbf{a},\mathbf{v})$ denote the $[n,k]_q$ GRS code with evaluation points $\mathbf{a} 
\in \F_q^n$ and factors $\mathbf{v} \in \F_q^n$ (see \autoref{sec:GRS}), and $\GRS_{k,r}(\mathbf{a},\mathbf{v})$ be the 
$[n,k]_q$ subcode of the $[n,k+1]_q$ GRS code $\GRS_{k+1}(\mathbf{a},\mathbf{v})$, 
whose generator matrix is derived by removing the row of degree $k-r$ from the 
generator matrix of $\GRS_{k+1}(\mathbf{a},\mathbf{v})$, where $1 \le r \le k-1$ (see \autoref{sec:SGRS}). 
In this paper, we will systematically study the self-dual property and the NMDS property of the codes 
$\GRS_{k,r}(\mathbf{a},\mathbf{v})$. For $r \in \setnd{1,k-1}$, equivalent characterizations for 
$\GRS_{k,r}(\mathbf{a},\mathbf{v})$ being self-dual are proposed (see \autoref{theorem:self_dual_1}, 
\autoref{theorem:self_dual_2}), while $\GRS_{k,r}(\mathbf{a},\mathbf{v})$ can never be self-dual 
when $2 \le r \le k-2$ (see \autoref{theorem:not_self_dual}). In particular, the equivalent characterization 
for $\GRS_{k,1}(\mathbf{a},\mathbf{v})$ being self-dual is closely related to that of the GRS codes 
being self-dual (compare \autoref{theorem:self_dual_1} and 
\autoref{theorem:GRS_self_dual}), and known constructions of self-dual GRS codes in the literature 
\cite{Jin2017SelfDualGRS,Yan2019SelfDualGRS,Fang2019SelfDualGRS,
	Fang2021SelfDualGRS,Ning2021SelfDualGRS,Wan2023SelfDualGRS}
can be exploited conveniently to the construction of self-dual $\GRS_{k,1}(\mathbf{a},\mathbf{v})$ 
(see \autoref{corollary:self_dual_from_GRS}). These results extended the known results on the 
self-dual property of $\GRS_{k,1}(\mathbf{a},\mathbf{v})$ in the literature \cite{Han2024NMDS}. 
On the other hand, equivalent conditions for $\GRS_{k,r}(\mathbf{a},\mathbf{v})$ being NMDS are proposed 
(see \autoref{theorem:nmds}). In particular, we show that $\GRS_{k,k-1}(\mathbf{a},\mathbf{v})$ is either NMDS or MDS, 
extending the results in \cite{Han2024NMDS}. The self-dual properties and NMDS properties of the 
subcodes of the extended GRS codes are also studied, extending the results in \cite{subcodeEGRS}. 
Finally, we study the dual codes of $\GRS_{k,r}(\mathbf{a},\mathbf{v})$ for $r \in \setnd{1,2}$. 
For $r = 1$, $\GRS_{k,1}(\mathbf{a},\mathbf{v})$ is closed under taking dual codes when the elements in $\mathbf{a}$ 
sum to zero (see \autoref{theorem:dual_code_1}), while the dual code of $\GRS_{k,1}(\mathbf{a},\mathbf{v})$ is a 
twisted GRS code when the elements in $\mathbf{a}$ do not sum to zero \cite{Huang2021twistedRS}. 
For $r = 2$, the dual code of $\GRS_{k,2}(\mathbf{a},\mathbf{v})$ is also related to the twisted GRS codes 
(see \autoref{theorem:dual_code_2}). The dual codes of the subcodes of the extended GRS codes are also studied 
for $r \in \setnd{1,2}$. 

This paper is organized as follows. \autoref{sec:preliminary} is 
devoted to preliminary knowledge, where we formally define the 
subcodes of the (extended) GRS codes considered in this paper, and 
introduce related known results. 
In \autoref{sec:self_dual_subcodes}, \autoref{sec:NMDS} and 
\autoref{sec:dual_code_subcodes}, we study the self-dual property, 
NMDS property and the dual codes of the subcodes of the (extended) GRS codes in the scope of this paper, 
respectively. Finally, \autoref{sec:conclusion} concludes this paper.

\section{Preliminary}
\label{sec:preliminary}
%We denote by $\F_q$ the finite field of $q$ elements, where $q$ is a 
%prime power. For positive integers $k \le n$, an $[n,k,d]_q$ linear 
%code $C$ is a linear subspace of $\F_q^n$ of dimension $k$ and minimum 
%Hamming distance $d$. The Singleton bound for linear codes states that 
%$d \le n-k+1$, and $[n,k,n-k+1]_q$ codes reaching the Singleton bound are 
%called \emph{maximum distance separable} (MDS) codes. 
%$[n,k,n-k]_q$ codes are called \emph{almost-MDS (AMDS)}. 
%Let $C$ be an $[n,k]_q$ code, the dual code of $C$ is defined as 
%$$
%\dual{C} = \set{\mathbf{w} \in \F_q^n}{\mathbf{w} \cdot \mathbf{c} = 0, 
%	\, \forall \mathbf{c} \in C},
%$$
%where $\mathbf{w} \cdot \mathbf{c} = \sum_{i=1}^nw_ic_i$ and 
%$\mathbf{w} = (w_1,\dots,w_n), \mathbf{c} = (c_1,\dots,c_n)$.
%It is clear that $\dual{C}$ is an $[n,n-k]_q$ code. If $C = \dual{C}$, then 
%$C$ is called self-dual. If both $C$ and $\dual{C}$ are AMDS codes, then $C$ is called 
%\emph{near-MDS (NMDS)}. 
In this section, we introduce preliminary knowledge and set up some notations used throughout the paper. 
\autoref{sec:GRS} is devoted to the GRS codes. In \autoref{sec:SGRS}, we define the class of subcodes of the 
GRS codes concerned in this paper, and introduce known related results. \autoref{sec:TGRS} is devoted to the twisted 
GRS codes, which will be the dual codes of the subcodes of the GRS codes in some cases. 
Finally, in \autoref{sec:formula}, we introduce some formulas crucial to the development of this paper.
\subsection{Generalized Reed-Solomon Codes}
\label{sec:GRS}
Generalized Reed-Solomon codes (GRS) are a class of widely known linear MDS codes. 
Let $k \le n$ be positive integers, and $q \ge n$ be a prime power.
Fix $n$ pairwise distinct evaluation points $a_1,\dots,a_n \in \F_q$, 
and $n$ non-zero factors $v_1,\dots,v_n \in \F_q^* = \F_q \setminus \setnd{0}$. 
The evaluation map on $\F_q[x]$ at the evaluation points 
$\mathbf{a} = (a_1,\dots,a_n)$ with factors $\mathbf{v} = (v_1,\dots,v_n)$ 
is defined as
\begin{equation}
\label{eqn:evmap}
\evmap_{\mathbf{a},\mathbf{v}} : 
\F_q[x] \to \F_q^n, \, f(x) \mapsto (v_1f(a_1),\dots,v_nf(a_n)),
\end{equation}
where $\F_q[x]$ denotes the vector space of polynomials in single variable $x$ with coefficients in $\F_q$. 
Note that $\evmap_{\mathbf{a},\mathbf{v}}$ is a linear map. 
Let $\F_q[x]_k$ denote the $k$-dimensional subspace of polynomials of 
degree less than $k$, which can spanned by 
$\setnd{1,x,x^2,\dots,x^{k-1}}$. The generalized Reed-Solomon (GRS) code
with evaluation points $\mathbf{a}$ and factors $\mathbf{v}$ is 
defined as the image of $\evmap_{\mathbf{a},\mathbf{v}}$ restricted to 
$\F_q[x]_k$, denoted as $\GRS_{k}(\mathbf{a},\mathbf{v}) \defeq  
\evmap_{\mathbf{a},\mathbf{v}}(\F_q[x]_k)$. 
It is known that $\GRS_k(\mathbf{a},\mathbf{v})$ is an 
$[n,k,n-k+1]_q$ MDS code \cite[Ch. 5]{CodeBook}.
%\cite[Theorem 5.3.1]{CodeBook}. 
Let 
\begin{equation}
	\label{eqn:Vandermonde}
	V(\mathbf{a}) = \spalignmat{
	a_1^0,a_2^0,\dots,a_n^0;
	a_1^1,a_2^1,\dots,a_n^1;
	\vdots,\vdots,\vdots,\vdots;
	a_1^{n-1},a_2^{n-1},\dots,a_{n}^{n-1}
	}
\end{equation}
be the Vandermonde matrix with evaluation points $\mathbf{a}$. It is well-known that $V(\mathbf{a})$ is invertible, as 
$a_1,\dots,a_n$ are pairwise distinct. 
%For integers $a \le b$, we write $[a,b] \defeq \setnd{a,a+1,\dots,b}$, and 
%$[n] \defeq [1,n]$. For $J \subset [0,n-1]$, let $V_J(\mathbf{a})$ be the 
For $J \subset [0,n-1]$, let $V_J(\mathbf{a})$ be the submatrix of $V(\mathbf{a})$ consisting of rows in $V(\mathbf{a})$ 
of degrees specified by $J$, i.e., 
\begin{equation}
	\label{eqn:Vandermonde_J}
	V_J(\mathbf{a}) \defeq (a_i^j)_{j \in J, i \in [n]},
\end{equation}
where $[n] \defeq \setnd{1,2,\dots,n}$. It is clear that $\GRS_k(\mathbf{a},\mathbf{v})$ has the following 
generator matrix
\begin{equation}
\label{eqn:GRSmatrix}
G_k(\mathbf{a},\mathbf{v}) \defeq V_{[0,k-1]}(\mathbf{a})D(\mathbf{v}),
\end{equation}
where $D(\mathbf{v})$ is the diagonal matrix with entries in $\mathbf{v}$.

Similarly, the extended generalized Reed-Solomon (EGRS) code with evaluation points $\mathbf{a}$ and factors $\mathbf{v}$ 
is defined as 
$$
\GRS_k(\mathbf{a},\mathbf{v},\infty) \defeq \set{
(v_1f(a_1),v_2f(a_2),\dots,v_nf(a_n),f[k-1])
}{f \in \F_q[x]_k},
$$
where $f[k-1]$ is the coefficient of the monomial $x^{k-1}$ in $f(x)$. 
It is well-known that $\GRS_k(\mathbf{a},\mathbf{v},\infty)$ is an $[n+1,k,n-k+2]_q$ MDS code \cite[Ch. 5]{CodeBook}. 
And $\GRS_k(\mathbf{a},\mathbf{v},\infty)$ has the following generator matrix 
\begin{equation}
	\label{eqn:EGRSmatrix}
G_k(\mathbf{a},\mathbf{v},\infty) \defeq  
\spalignmat{
a_1^0,\dots,a_n^0,0;
a_1^1,\dots,a_n^1,0;
\vdots,\vdots,\vdots,\vdots;
a_1^{k-2},\dots,a_n^{k-2},0;
a_1^{k-1},\dots,a_n^{k-1},1
}D(\mathbf{v},1).
\end{equation}

There were many works on the construction of 
self-dual (extended) GRS codes, for which we refer to 
\cite{Jin2017SelfDualGRS,Yan2019SelfDualGRS,Fang2019SelfDualGRS,
	Fang2021SelfDualGRS,Ning2021SelfDualGRS,Wan2023SelfDualGRS} and the 
references therein. In particular, we have the following characterization of when 
%(extended) 
GRS codes are self-dual \cite{Yan2019SelfDualGRS,Ning2021SelfDualGRS}. 
\begin{theorem}[\cite{Yan2019SelfDualGRS,Ning2021SelfDualGRS}]
	\label{theorem:GRS_self_dual}
Let $n = 2k \le q$, $\mathbf{a} = (a_1,\dots,a_n)$ be $n$ pairwise distinct 
evaluation points in $\F_q$ and $u_i = \prod_{j \in [n], j \neq i}(a_i-a_j)^{-1}, i \in [n]$. 
Then, there exist non-zero factors $v \in (\F_q^*)^n$ 
such that $\GRS_{k}(\mathbf{a},\mathbf{v})$ is self-dual if and only if 
$u_1,\dots,u_n$ are simultaneously squares or non-squares in $\F_q$. 
\end{theorem}
%\begin{theorem}[\cite{Yan2019SelfDualGRS,Ning2021SelfDualGRS}]
%Let $n+1 = 2k$ with $n \le q$, $\mathbf{a} = (a_1,\dots,a_n)$ be $n$ pairwise distinct 
%evaluation points in $\F_q$ and $u_i = \prod_{j \in [n], j \neq i}(a_i-a_j)^{-1}, i \in [n]$. 
%Then, there exists $\mathbf{v} \in (\F_q^*)^n$ such that $\GRS_k(\mathbf{a},\mathbf{v},\infty)$ is 
%self-dual if and only if $-u_1,\dots,-u_n$ are all squares in $\F_q$.
%\end{theorem}

\subsection{Subcodes of Generalized Reed-Solomon Codes}
\label{sec:SGRS}
Let $J(k,r) \defeq [0,k] \setminus \setnd{k-r}$ for $r \in [1,k-1]$, and 
$\F_q[x]_{k,r} \defeq \spn \set{x^j}{j \in J(k,r)}$. 
Note that 
$$
\F_q[x]_{k,r} = \set{f \in \F_q[x]_{k+1}}{f[k-r] = 0}
$$
consists of polynomials of degree at most $k$ not containing monomials of degree $k-r$. 
Let $\mathbf{a} = (a_1,\dots,a_n)$ be pairwise distinct evaluation points in $\F_q$ and 
$\mathbf{v} = (v_1,\dots,v_n) \in (\F_q^*)^n$ be non-zero factors. 
For $r \in [1,k-1]$ and $k \le n-1$, 
\begin{equation}
\GRS_{k,r}(\mathbf{a},\mathbf{v}) \defeq \evmap_{\mathbf{a},\mathbf{v}}(\F_q[x]_{k,r})
\end{equation}
is a subcode of $\GRS_{k+1}(\mathbf{a},\mathbf{v})$, which will be concerned in this paper. 
It is clear that $\GRS_{k,r}(\mathbf{a},\mathbf{v})$ is an $[n,k]_q$ code with 
the following generator matrix 
\begin{equation}
	\label{eqn:matrix_subcode_GRS}
G_{k,r}(\mathbf{a},\mathbf{v}) \defeq V_{J(k,r)}(\mathbf{a})D(\mathbf{v}) = 
\spalignmat[c]{
	v_1a_1^0,\dots,v_na_n^0;
	\vdots,\vdots,\vdots;
	v_1a_1^{k-r-1},\dots,v_na_n^{k-r-1};
	v_1a_1^{k-r+1},\dots,v_na_n^{k-r+1};
	\vdots,\vdots,\vdots;
	v_1a_1^k,\dots,v_na_n^k
}.
\end{equation}
It is known that $\GRS_{k,r}(\mathbf{a},\mathbf{v})$ is either AMDS or non-GRS MDS \cite{Jin2024nonRSMDS}. 
And $\GRS_{k,r}(\mathbf{a},\mathbf{v})$ is MDS if and only if the evaluation of  
$s_{k,r}$ is not zero on any $k$-subset of $\mathbf{a}$ \cite{Jin2024nonRSMDS}, 
where 
\begin{equation}
\label{eqn:elementary_symmetric_polynomial}
s_{k,r}(X_1,\dots,X_k) \defeq \sum_{1 \le j_1 < \dots < j_r \le k}
X_{j_1}X_{j_2}\dots X_{j_r}
\end{equation}
is the elementary symmetric polynomial in $k$ variables of degree $r$. 
For simplicity of notations, we also write 
\begin{equation}
\label{eqn:elementary_symmetric_polynomial_simplified}
s_r(X_1,\dots,X_k) \defeq s_{k,r}(X_1,\dots,X_k),
\end{equation}
when the number of variables $k$ is clear from the context. 
When $r = 1$, $\GRS_{k,1}(\mathbf{a},\mathbf{v})$ is either NMDS or MDS \cite{Han2024NMDS}, 
and self-dual codes were constructed out of $\GRS_{k,1}(\mathbf{a},\mathbf{v})$ \cite{Han2024NMDS}. 
If $s_{1}(\mathbf{a}) \neq 0$, the dual code of $\GRS_{k,1}(\mathbf{a},\mathbf{v})$ is a twisted GRS code 
\cite{Huang2021twistedRS}. For $r \ge 2$, the corresponding properties of $\GRS_{k,r}(\mathbf{a},\mathbf{v})$ 
have not been considered in the literature to the best of our knowledge, 
and will be studied in this paper. 

Similarly, for $r \in [1,k-1]$ and $k \le n-1$, we also consider the following subcodes of the extended GRS codes:
\begin{equation}
\GRS_{k,r}(\mathbf{a},\mathbf{v},\infty) = \set{(v_1f(a_1),v_2f(a_2),\dots,v_nf(a_n),f[k])}{f \in \F_q[x]_{k,r}}.
\end{equation}
Clearly, $\GRS_{k,r}(\mathbf{a},\mathbf{v},\infty)$ is an $[n+1,k]_q$ code. 
When $r = 1$, $\GRS_{k,1}(\mathbf{a},\mathbf{v},\infty)$ has been thoroughly 
studied in \cite{subcodeEGRS}. In particular, 
$\GRS_{k,1}(\mathbf{a},\mathbf{v},\infty)$ is either NMDS or non-GRS MDS  
\cite{subcodeEGRS}, and $\GRS_{k,1}(\mathbf{a},\mathbf{v},\infty)$ is 
MDS if and only if $s_{k,1}$ is not zero on any $k$-subset of $\mathbf{a}$ 
\cite{subcodeEGRS}. When $n+1 = 2k$, 
$\GRS_{k,1}(\mathbf{a},\mathbf{v},\infty)$ can never be self-dual for any 
choice of evaluation points $\mathbf{a}$ and factors $\mathbf{v}$ 
\cite{subcodeEGRS}. It is clear that $\GRS_{k,r}(\mathbf{a},\mathbf{v},\infty)$ 
has the following generator matrix 
\begin{equation}
	\label{eqn:matrix_subcode_EGRS}
G_{k,r}(\mathbf{a},\mathbf{v},\infty) \defeq 
\spalignmat[c]{
	v_1a_1^0,\dots,v_na_n^0,0;
	\vdots,\vdots,\vdots,\vdots,;
	v_1a_1^{k-r-1},\dots,v_na_n^{k-r-1},0;
	v_1a_1^{k-r+1},\dots,v_na_n^{k-r+1},0;
	\vdots,\vdots,\vdots,\vdots;
	v_1a_1^k,\dots,v_na_n^k,1
}.	
\end{equation}
It was also shown in \cite{subcodeEGRS} that 
$\GRS_{k,1}(\mathbf{a},\mathbf{v},\infty)$ 
has the following parity check matrix
\begin{equation}
	\label{eqn:subcode1_EGRS_dual}
\spalignmat[c]{
	a_1^0,\dots,a_n^0,0;
	a_1^{1},\dots,a_n^{1},0;
	\vdots,\vdots,\vdots,\vdots;
	a_1^{n-k-2},\dots,a_n^{n-k-2},0;
	a_1^{n-k-1},\dots,a_n^{n-k-1},-1;
	a_1^{n-k},\dots,a_n^{n-k},-\sum_{i=1}^na_i
}\spalignmat{
	\frac{u_1}{v_1};
	,\frac{u_2}{v_2};
	,,\ddots;
	,,,\frac{u_n}{v_n};
	,,,,1
}.	
\end{equation}
For $r \ge 2$, the corresponding properties of $\GRS_{k,r}(\mathbf{a},\mathbf{v},\infty)$ have not 
been considered in the literature, to the best of our knowledge. 

For $\mathbf{c} = (c_1,\dots,c_n) \in \F_q^n$, we denote the Hamming weight of $\mathbf{c}$ as $\wt(\mathbf{c})$, i.e., 
\begin{equation}
\wt(\mathbf{c}) = \card{\set{i \in [n]}{c_i \neq 0}}.
\end{equation}
Similar as in \cite{Jin2024nonRSMDS,subcodeEGRS}, we characterize the minimum distance of 
$\GRS_{k,r}(\mathbf{a},\mathbf{v},\infty)$ in the following lemma. 
\begin{lemma}
	\label{lemma:subcode_EGRS_MDS}
Let $1 \le k \le n-1$, $n \le q$, $r \in [1,k-1]$, $\mathbf{a} = (a_1,\dots,a_n)$ 
be pairwise distinct evaluation points in $\F_q$, and 
$\mathbf{v}=(v_1,\dots,v_n) \in (\F_q^*)^n$ be non-zero factors. 
Then, $\GRS_{k,r}(\mathbf{a},\mathbf{v},\infty)$ is AMDS or MDS. 
Moreover, $\GRS_{k,r}(\mathbf{a},\mathbf{v},\infty)$ is not MDS if and only if 
there is a subset $Z_1 \subset \mathbf{a}$ of size $k-1$ with 
$s_{r-1}(Z_1) = 0$, or there is a subset 
$Z_2 \subset \mathbf{a}$ of size $k$ with $s_{r}(Z_2) = 0$.
\end{lemma}
\begin{proof}
Any codeword of $\GRS_{k,r}(\mathbf{a},\mathbf{v},\infty)$ is of the form 
$\mathbf{c} = (v_1f(a_1),\dots,v_nf(a_n),f[k])$ for some polynomial 
$f(x) \in \F_q[x]_{k,r}$. If $f[k] = 0$, then 
$\deg f(x) \le k-1$, and $f(x)$ has at most $k-1$ zeros, and thus 
$\wt(\mathbf{c}) \ge n-k+1$. On the other hand, if $f[k] \neq 0$, then 
$f(x)$ has at most $k$ zeros, and $\wt(\mathbf{c}) \ge n-k+1$. 
So the minimum distance of $\GRS_{k,r}(\mathbf{a},\mathbf{v},\infty)$ is 
at least $n-k+1$, and $\GRS_{k,r}(\mathbf{a},\mathbf{v},\infty)$ is 
AMDS or MDS.

Suppose that $\GRS_{k,r}(\mathbf{a},\mathbf{v},\infty)$ is not MDS.	
Then, there is a codeword $\mathbf{c} \in 
\GRS_{k,r}(\mathbf{a},\mathbf{v},\infty)$ of weight $n-k+1$. 
By definition, $\mathbf{c} = (v_1f(a_1),\dots,v_nf(a_n),f[k])$ for 
some polynomial $f(x) \in \F_q[x]_{k,r}$. After a scaling if necessary, we 
may also assume that $f(x)$ is monic. If $f[k] = 0$, $f(x)$ has 
$k-1$ zeros among $\mathbf{a}$. Let $Z_1 \subset \mathbf{a}$ 
be the set of zeros of $f(x)$. As $f(x) \neq 0$ and $\deg f(x) \le k-1$, 
we know that $f(x) = f_{Z_1}(x) \defeq \prod_{z \in Z_1}(x-z)$. 
As $f(x)$ contains no monomials of degree $k-r$, we have that 
$s_{r-1}(Z_1) = 0$. On the other hand, if $f[k] \neq 0$, $f(x)$ has 
$k$ zeros among $\mathbf{a}$. Again, let $Z_2 \subset \mathbf{a}$ be the 
zeros of $f(x)$, then $f(x) = f_{Z_2}(x)$. As $f(x) \in \F_q[x]_{k,r}$, we have that 
$s_{r}(Z_2) = 0$. In conclusion, if $\GRS_{k,r}(\mathbf{a},\mathbf{v},\infty)$ is not MDS, then 
there is a $(k-1)$-subset $Z_1$ of $\mathbf{a}$ such that $s_{r-1}(Z_1) = 0$, or there is a 
$k$-subset $Z_2$ of $\mathbf{a}$ such that $s_{r}(Z_2) = 0$.

On the contrary, if $Z_1 \subset \mathbf{a}$ is a $(k-1)$-subset such that 
$s_{r-1}(Z_1) = 0$, then $f_{Z_1}(x) \in \F_q[x]_{k,r}$ and 
$(v_1f_{Z_1}(a_1),\dots,v_nf_{Z_1}(a_n),f_{Z_1}[k])$ is a 
codeword in $\GRS_{k,r}(\mathbf{a},\mathbf{v},\infty)$ of weight $n-k+1$. 
Similarly, if $Z_2 \subset \mathbf{a}$ is a $k$-subset such that 
$s_{r}(Z_2) = 0$, then $(v_1f_{Z_2}(a_1),\dots,
v_nf_{Z_2}(a_n),f_{Z_2}[k]) \in \GRS_{k,r}(\mathbf{a},\mathbf{v},\infty)$ is a 
codeword of weight $n-k+1$. In both cases, $\GRS_{k,r}(\mathbf{a},\mathbf{v},
\infty)$ is not MDS.
\end{proof}
For $r = 1$, we have $s_{k-1,r-1} = 1$. Then, by 
\autoref{lemma:subcode_EGRS_MDS}, $\GRS_{k,1}(\mathbf{a},\mathbf{v},\infty)$ 
is MDS if and only if $s_{k,1}(Z_2) \neq 0$ for any $k$-subset 
$Z_2 \subset \mathbf{a}$, which recovers the 
characterization of $\GRS_{k,1}(\mathbf{a},\mathbf{v},\infty)$ being 
MDS in \cite{subcodeEGRS}.

\subsection{Twisted Generalized Reed-Solomon Codes}
\label{sec:TGRS}
In this section, we review the twisted generalized Reed-Solomon codes 
introduced in 
\cite{Beelen2017ISITtwistedRS,Beelen2018ISITtwistedRS,Beelen2022ITtwistedRS}. 
Twisted GRS codes are not mainly concerned in this paper, but they appear as 
dual codes of the subcodes of the GRS codes in some cases. 
We have the following linear isomorphism:
\begin{equation}
	\label{eqn:phi}
	\phi : \F_q^n \to \F_q[x]_n, \, 
	(f_0,\dots,f_{n-1}) \mapsto \sum_{i=0}^{n-1} f_ix^i.
\end{equation}
Let 
$$
\bm{\eta} = \spalignmat{
\bm{\eta}_0; \vdots; \bm{\eta}_{k-1}
}
= (\eta_{i,j})_{i \in [0,k-1],j \in [0,n-k-1]}
$$
be a $k \times (n-k)$ matrix over $\F_q$, where 
$\bm{\eta}_0,\dots,\bm{\eta}_{k-1}$ are the rows of 
$\bm{\eta}$, and $\eta_{i,j}$ are the entries. Let 
$$
G(\eta) = \spalignmat{I_k, \bm{\eta}}
= \spalignmat{
\mathbf{g}_0;\vdots;\mathbf{g}_{k-1}
},
$$
where $I_k$ is the $k \times k$ identity matrix, and $\mathbf{g}_i$ are 
the rows of $G(\bm{\eta})$. Let 
$$
g_i(x) = \phi(\mathbf{g}_i) = 
x^i + \sum_{j=0}^{n-k-1}\eta_{i,j}x^{k+j}, \quad i \in [0,k-1],
$$
and $W(\bm{\eta}) = \spn \setnd{g_0(x),\dots,g_{k-1}(x)}$. 
In its most general form, the twisted GRS (TGRS) code with evaluation 
points $\mathbf{a}$ and factors $\mathbf{v}$ is defined as the image of 
$\evmap_{\mathbf{a},\mathbf{v}}$ restricted on $W(\bm{\eta})$, i.e., 
$\evmap_{\mathbf{a},\mathbf{v}}(W(\bm{\eta}))$.

In some cases, TGRS codes are known to be dual codes of subcodes of GRS 
codes. For example, let $\eta \in \F_q^*$ and 
$$W_{+,k,\eta} \defeq \spn \setnd{1,x,x^2,\dots,x^{k-2},x^{k-1} + \eta x^k}.$$
The $(+)$-TGRS code with evaluation points $\mathbf{a}$ and 
factors $\mathbf{v}$ is defined as $C_{+,k,\eta}(\mathbf{a},\mathbf{v}) \defeq  
\evmap_{\mathbf{a},\mathbf{v}}(W_{+,k,\eta})$ \cite{Beelen2022ITtwistedRS}. 
The dual code of $C_{+,k,\eta}(\mathbf{a},\mathbf{v})$ was 
studied in \cite{Huang2021twistedRS}. In particular, 
if $s_{n,1}(\mathbf{a}) \neq 0$ and $\eta = -1/s_{n,1}(\mathbf{a})$, then 
\begin{equation}
	\label{eqn:subcode1_GRS_dual_t1_nonzero}
	\dual{C_{+,k,\eta}(\mathbf{a},\mathbf{v})} = 
	\GRS_{n-k,1}\qty(\mathbf{a},(u_i/v_i)_{i \in [n]})
\end{equation}
is the subcode of the GRS codes in the scope of this paper, where $u_i = \prod_{j \in [n], j \neq i}(a_i-a_j)^{-1}$. 
It is also known that $C_{+,k,\eta}(\mathbf{a},\mathbf{v})$ is MDS if 
and only if any $k$-subset of $\mathbf{a}$ does not sum to $-\eta^{-1}$ 
\cite{Beelen2022ITtwistedRS}. Moreover, if $s_1(\mathbf{a}) \neq 0$ and $n = 2k$, then both 
$C_{+,k,\eta}(\mathbf{a},\mathbf{v})$ and 
$\GRS_{k,1}\qty(\mathbf{a},\mathbf{v})$ 
cannot be self-dual for any choice of factors $\mathbf{v} \in (\F_q^*)^n$ 
\cite{Huang2021twistedRS}. We note that $C_{+,k,\eta}(\mathbf{a},\mathbf{v})$ 
has the following generator matrix 
\begin{equation}
\label{eqn:matrix_PTGRS}
G_{+,k,\eta}(\mathbf{a},\mathbf{v}) \defeq 
\spalignmat[c]{
a_1^0, \dots, a_n^0 ;
\vdots, \vdots, \vdots;
a_1^{k-2}, \dots, a_n^{k-2};
a_1^{k-1}+\eta a_1^k,\dots,a_n^{k-1}+\eta a_n^k
}D(\mathbf{v}).
\end{equation}
\subsection{Two Formulas}
\label{sec:formula}
The following formula in \autoref{lemma:sum} appeared in 
\cite{Jin2017SelfDualGRS,Ning2021SelfDualGRS,Huang2021twistedRS,subcodeGRSVariant}
and will be crucial to the development of this paper. We provide with a proof from the view point of 
Lagrange's interpolation theorem, which is suitable for the generalization of the formula. 
\begin{lemma}[\cite{Jin2017SelfDualGRS,Ning2021SelfDualGRS}]
	\label{lemma:sum}
	Let $\mathbf{a} = (a_1,\dots,a_n)$ be pairwise distinct elements in $\F_q$, and 
	$u_i \defeq \prod_{j=1, j \neq i}^n (a_j-a_i)^{-1}, i \in [n]$. Then, 
	\begin{equation}
		\label{eqn:dual}
		\sum_{i=1}^n a_i^l u_i = 
		\begin{cases}
			0, & l \in [0,n-2], \\
			1, & l = n-1, \\
			t_1, & l = n, \\
			t_1^2-t_2, & l = n+1.
		\end{cases}
	\end{equation}
	where $t_i \defeq s_{i}(\mathbf{a})$ and $s_{i}(X_1,\dots,X_n)$ is the elementary symmetric polynomials defined 
	in \autoref{eqn:elementary_symmetric_polynomial} and \autoref{eqn:elementary_symmetric_polynomial_simplified}. 
%	For $l \ge n+2$, $\sum_{i=1}^n u_ia_i^l$ can be written as a 
%	polynomial in $t_1,\dots,t_n$. 
\end{lemma}
\begin{proof}
	By Lagrange's interpolation theorem \cite[Ch. 1]{FiniteFields}, we 
	know that 
	\begin{equation}
		\label{eqn:Lagrange}
		\sum_{i=1}^n a_i^l u_i \prod_{j \in [n], j \neq i}(x-a_j) = x^l 
		\pmod{f_{\mathbf{a}}(x)}, 
	\end{equation}
	where $f_{\mathbf{a}}(x) \defeq \prod_{i=1}^n (x-a_i)$. 
	Comparing the coefficients of $x^{n-1}$, we have that 
	$$
	\sum_{i=1}^n a_i^l u_i = 
	\begin{cases}
		0, & l \in [0,n-2], \\
		1, & l = n-1. \\
	\end{cases}
	$$
	
	For $l = n$, we have 
	\begin{align}
		x^n \pmod{f_\mathbf{a}(x)} &= x^n - f_\mathbf{a}(x) \\
		&= t_1x^{n-1} - t_2x^{n-2} + 
		t_3x^{n-3} + \dots + (-1)^{n+1}t_n.	
		\label{eqn:ed1}
	\end{align}
	Comparing the coefficients of $x^{n-1}$, we have that 
	$$
	\sum_{i=1}^n u_ia_i^n = t_1.
	$$
	
	Finally, we consider the case $l = n+1$. 
	By the linear isomorphism in \autoref{eqn:phi}, 
	we rewrite \autoref{eqn:ed1} as 
	$$
	x^n \pmod{f_\mathbf{a}(x)} \leftrightarrow 
	(t_1,-t_2,t_3,\dots,(-1)^{n+1}t_n). 
	$$
	Then 
	\begin{align*}
		x^{n+1} \pmod{f_\mathbf{a}(x)} \leftrightarrow &
		(-t_2,t_3,\dots,(-1)^{n+1}t_n,0) + 
		t_1(t_1,-t_2,t_3,\dots,(-1)^{n+1}t_n) \\
		&=(t_1^2-t_2,-t_1t_2-t_3,\dots,).	
	\end{align*}
	We conclude that 
	$$
	\sum_{i=1}^n u_ia_i^{n+1} = t_1^2-t_2.
	$$
%	By an induction, we know that $\sum_{i=1}^n u_ia_i^l$ is a 
%	polynomial in $t_1,\dots,t_n$ for $l \ge n+2$.
\end{proof}
Similarly, by comparing the coefficients of $x$ in \autoref{eqn:Lagrange}, 
we have the following formula in \autoref{lemma:sum2}. 
\begin{lemma}
	\label{lemma:sum2}
Let $\mathbf{a} = (a_1,\dots,a_n)$ be pairwise distinct elements in 
$\F_q$, and $\mathbf{a}_i \defeq \mathbf{a} \setminus \setnd{a_i}$ for $i \in [n]$. Then,
	$$
	\sum_{i=1}^na_i^lu_is_{n-2}(\mathbf{a}_i) 
	= 
	\begin{cases}
		(-1)^n, & l = 1, \\
		0, & l \in \setnd{0} \cup [2,n-1], \\
		t_{n-1}, & l = n, \\
		t_1t_{n-1}-t_n, & l = n+1.
	\end{cases}
	$$
\end{lemma}
\autoref{lemma:sum2} can be proved by comparing the coefficient of $x$ in \autoref{eqn:Lagrange} in a 
similar way to the proof of \autoref{lemma:sum}. The details are omitted.

\section{Self-Dual Subcodes of GRS Codes}
\label{sec:self_dual_subcodes}
In this section, we determine when $\GRS_{k,r}(\mathbf{a},\mathbf{v})$ and 
$\GRS_{k,r}(\mathbf{a},\mathbf{v},\infty)$ are self-dual. Note that self-dual codes must be of 
parameters $[2k,k]_q$. In \autoref{sec:SelfDual2}, we show that 
$\GRS_{k,r}(\mathbf{a},\mathbf{v})$ or $\GRS_{k,r}(\mathbf{a},\mathbf{v})$ 
can never be self-dual for any evaluation points $\mathbf{a}$ and factors $\mathbf{v}$ when $r \in [2,k-2]$. 
%(see \autoref{theorem:not_self_dual} and \autoref{theorem:not_self_dual_E}). 
Then, \autoref{sec:SelfDual1} is devoted to the characterization of when $\GRS_{k,r}(\mathbf{a},\mathbf{v})$ or 
$\GRS_{k,r}(\mathbf{a},\mathbf{v},\infty)$ is self-dual for $r \in \setnd{1,k-1}$. 
For $r = 1$, there were already some results in the literature. In particular, 
it was shown in \cite{subcodeEGRS} that $\GRS_{k,1}(\mathbf{a},\mathbf{v},\infty)$ 
can never be self-dual, and a sufficient condition for $\GRS_{k,1}(\mathbf{a},\mathbf{v})$ being self-dual was 
proposed in \cite{Han2024NMDS}.

Let $J_1,J_2$ be two set of integers, and we define $J_1+J_2 \defeq \set{j_1+j_2}{j_1 \in J_1, j_2 \in J_2}$. 
For $r \in [1,k-1]$, recall that $J(k,r) \defeq [0,k] \setminus \setnd{k-r}$. 
When $k \ge 4$, it is easy to check that 
\begin{equation}
	\label{eqn:2J}
	J(k,r) + J(k,r) = 
	\begin{cases}
		J(2k,1), & r = 1, \\ 
		J(2k,2k-1), & r = k-1 , \\
		[0,2k], & r \in [2,k-2].
	\end{cases}
\end{equation}
Now we are ready to determine when the subcodes of the GRS codes are self-dual. 
\subsection{The case $r \in [2,k-2]$}
\label{sec:SelfDual2}
First, we show that $\GRS_{k,r}(\mathbf{a},\mathbf{v})$ and $\GRS_{k,r}(\mathbf{a},\mathbf{v},\infty)$ 
can never be self-dual for any evaluation points $\mathbf{a}$ and factors $\mathbf{v}$ when $r \in [2,k-2]$.
\begin{theorem}
	\label{theorem:not_self_dual}
	Let $n = 2k \ge 8$ and $r \in [2,k-2]$. Then for any pairwise distinct 
	evaluation points 
	$\mathbf{a} = (a_1,\dots,a_n)$ and non-zero factors $\mathbf{v} = 
	(v_1,\dots,v_n)$ in $\F_q$, 
	$\GRS_{k,r}(\mathbf{a},\mathbf{v})$ is not self-dual.
\end{theorem}
\begin{proof}
	Let $G = G_{k,r}(\mathbf{a},\mathbf{v})$ be the generator matrix of  
	$\GRS_{k,r}(\mathbf{a},\mathbf{v})$ defined in 
	\autoref{eqn:matrix_subcode_GRS}. Note that 
	$\GRS_{k,r}(\mathbf{a},\mathbf{v})$ is self-dual if and only if 
	$GG^\intercal = 0$. By calculation, entries of $GG^\intercal$ are of the 
	form 
	$$
	\sum_{l = 1}^n v_l^2a_l^{i+j}, \, i,j \in J(k,r).
	$$
	By \autoref{eqn:2J}, the set of entries in $GG^\intercal$ equals 
	$
	\set{\sum_{l = 1}^n v_l^2a_l^i}{i \in [0,2k]}.
	$
	Thus, $GG^\intercal = 0$ is equivalent to that 
%	\begin{equation}
%		\label{eqn:not_self_dual_1}
%		\sum_{l \in [n]}v_l^2a_l^i = 0, \, \forall i \in [0,2k].
%	\end{equation}
%	\autoref{eqn:not_self_dual_1} can be rewritten in terms of matrices as 
	\begin{equation}
		\label{eqn:not_self_dual_2}
		\spalignmat{
			a_1^0,\dots,a_n^0;
			\vdots,\vdots,\vdots;
			a_1^n,\dots,a_n^n
		}
		\spalignmat{
			v_1^2;\vdots;v_n^2
		} = 0.
	\end{equation}
	But \autoref{eqn:not_self_dual_2} is impossible, as 
	the coefficient matrix contains the invertible Vandermonde matrix 
	$V(\mathbf{a})$ (see \autoref{eqn:Vandermonde}) as a submatrix, and 
	$v_1^2,\dots,v_n^2$ are non-zero.
	We conclude that $\GRS_{k,r}(\mathbf{a},\mathbf{v})$ is not self-dual.
\end{proof}
\begin{theorem}
	\label{theorem:not_self_dual_E}
Let $n+1 = 2k \ge 8$ and $r \in [2,k-2]$. Then for any pairwise distinct 
evaluation points $\mathbf{a} = (a_1,\dots,a_n)$ and non-zero factors 
$\mathbf{v} = (v_1,\dots,v_n)$ in $\F_q$, 
$\GRS_{k,r}(\mathbf{a},\mathbf{v},\infty)$ is not self-dual.
\end{theorem}
The proof of \autoref{theorem:not_self_dual_E} is similar to that of \autoref{theorem:not_self_dual}, and 
details are omitted.
%\begin{proof}
%The proof is similar to that of \autoref{theorem:not_self_dual}. 
%Let $G = G_{k,r}(\mathbf{a},\mathbf{v},\infty)$ be the generator matrix for $\GRS_{k,r}(\mathbf{a},\mathbf{v},\infty)$ 
%in \autoref{eqn:matrix_subcode_EGRS}. Then, $\GRS_{k,r}(\mathbf{a},\mathbf{v},\infty)$ is self-dual if and only 
%if $GG^\intercal = 0$, which is equivalent to that 
%\begin{equation}
%	\label{eqn:not_self_dual_E}
%V_{[0,2k-1]}(\mathbf{a})(v_1^2,\dots,v_n^2)^\intercal = 0 
%\qand \sum_{i=1}^nv_i^2a_i^{2k} + 1 = 0.
%\end{equation}
%But \autoref{eqn:not_self_dual_E} is impossible, as $V_{[0,2k-1]}(\mathbf{a})$ contains the Vandermonde 
%matrix $V(\mathbf{a})$ as a submatrix and $v_1^2,\dots,v_n^2$ are non-zero.
%\end{proof}

\subsection{The case $r \in \setnd{1,k-1}$}
\label{sec:SelfDual1}
In this section, we characterize when $\GRS_{k,r}(\mathbf{a},\mathbf{v})$ or 
$\GRS_{k,r}(\mathbf{a},\mathbf{v},\infty)$ is self-dual for $r \in 
\setnd{1,k-1}$. For $r = 1$, it was known that $\GRS_{k,1}(\mathbf{a},\mathbf{v},\infty)$ can never be 
self-dual \cite{subcodeEGRS}. So, we only consider the code $\GRS_{k,1}(\mathbf{a},\mathbf{v})$ in 
\autoref{theorem:self_dual_1}. We note that the sufficiency part of \autoref{theorem:self_dual_1} was already 
known in \cite{Han2024NMDS}. 
\begin{theorem}
	\label{theorem:self_dual_1}
	Let $n = 2k \ge 8$, and $\mathbf{a} = (a_1,\dots,a_n)$ be pairwise distinct 
	evaluation points in $\F_q$. 
	Then, there exist non-zero factors $\mathbf{v} = (v_1,\dots,v_n) \in 
	(\F_q^*)^n$ such that 
	$\GRS_{k,1}(\mathbf{a},\mathbf{v})$ is self-dual if and only if 
	$t_1 = 0$ and $u_1,\dots,u_n$ are simultaneously squares or non-squares in 
	$\F_q$, where 
	$t_1,u_1,\dots,u_n$ are defined as in \autoref{lemma:sum}.
\end{theorem}
\begin{proof}
	$\implies$: Let $\mathbf{v} = (v_1,\dots,v_n) \in (\F_q^*)^n$ be $n$ 
	non-zero factors, and assume that 
	$\GRS_{k,1}(\mathbf{a},\mathbf{v})$ is self-dual. 
	Let $G$ be the generator matrix for $\GRS_{k,1}(\mathbf{a},\mathbf{v})$ 
	in \autoref{eqn:matrix_subcode_GRS}, then $GG^\intercal = 0$. Similar as in the proof of 
	\autoref{theorem:not_self_dual}, 
	the set of entries in $GG^\intercal$ equals 
	$$
	\set{\sum_{l \in [n]}v_l^2a_l^i}{i \in J(2k,1) = J(k,1) + J(k,1)}.
	$$
	Let 
	$$
	M = \spalignmat{
		a_1^0,\dots,a_n^0;
		\vdots,\vdots,\vdots;
		a_1^{2k-2},\dots,a_n^{2k-2};
		a_1^{2k},\dots,a_n^{2k}
	}
	$$
	and $N$ denote the submatrix of $M$ consisting of the first $2k-1$ rows. 
	Then, 
	$GG^\intercal = 0$ is equivalent to $M\mathbf{v}_2^\intercal = 0$, where 
	$\mathbf{v}_2 = (v_1^2,\dots,v_n^2)$. Note that $N = 
	V_{[0,2k-2]}(\mathbf{a})$ is a submatrix of the Vandermonde matrix (see \autoref{eqn:Vandermonde_J}). 
	Therefore, the kernel of $N$ has dimension $1$, and is spanned by 
	$\mathbf{v}_2$. By \autoref{lemma:sum}, $\mathbf{u} = (u_1,\dots,u_n)$ also 
	lives in the kernel of $N$ and is non-zero. We conclude that there is 
	$\lambda 
	\in \F_q^*$ such that $\mathbf{u} = \lambda \mathbf{v}_2$. In particular, 
	we 
	have shown that $u_1,\dots,u_n$ are simultaneouly squares (when $\lambda$ 
	is a 
	square) or non-squares (when $\lambda$ is a non-square) in $\F_q$. On the 
	other 
	hand, 
	$$
	0 = (a_1^{2k},\dots,a_n^{2k})\mathbf{v}_2^\intercal = \frac{1}{\lambda} 
	(a_1^{2k},\dots,a_{n}^{2k})
	\mathbf{u}^\intercal = \frac{t_1}{\lambda}.
	$$
	Thus, $t_1 = 0$.
	
	$\impliedby$: Conversely, assume that $t_1 = 0$ and $u_1,\dots,u_n$ are all 
	squares or non-squares. 
	Take $\lambda \in \F_q^*$ such that $\lambda u_1,\dots,\lambda u_n$ are all 
	squares. By \autoref{lemma:sum}, 
	$\GRS_{k,1}(\mathbf{a},(\sqrt{\lambda u_i})_{i \in [n]})$ is self-dual.
\end{proof}
\autoref{theorem:self_dual_1} can be thought of as an extension of 
\autoref{theorem:GRS_self_dual}. There were many works on the 
explicit construction of self-dual GRS codes by 
\autoref{theorem:GRS_self_dual} 
\cite{Jin2017SelfDualGRS,Yan2019SelfDualGRS,Fang2019SelfDualGRS,
	Fang2021SelfDualGRS,Ning2021SelfDualGRS,Wan2023SelfDualGRS}. 
These results can also be exploited conveniently for the explict 
construction of self-dual $\GRS_{k,1}(\mathbf{a},\mathbf{v})$
as in the following corollary. Recall that $\GRS_{k,1}(\mathbf{a},\mathbf{v})$ is always 
NMDS or MDS \cite{Han2024NMDS}, so the resulting codes are self-dual NMDS or self-dual MDS.
\begin{corollary}
	\label{corollary:self_dual_from_GRS}
Let $n = 2k$ be even and coprime to $q$. 
If $\GRS_{k}(\mathbf{a},\mathbf{v})$ is self-dual for some 
evaluation points $\mathbf{a} = (a_1,\dots,a_n)$ and factors 
$\mathbf{v} \in (\F_q^*)^n$, then there is $\mathbf{w} \in (\F_q^*)^n$ such 
that $\GRS_{k,1}(\mathbf{b},\mathbf{w})$ is self-dual, 
where $\mathbf{b} = (a_1-t_1/n,\dots,a_n-t_1/n)$, and $t_1 = s_{n,1}(\mathbf{a})$.
\end{corollary}
\begin{proof}
	As $n$ and $q$ are coprime, $1/n$ is well-defined in $\F_q$. 
	Write $\mathbf{b} \defeq (b_1,\dots,b_n)$, i.e., $b_i \defeq a_i-t_1/n$ are the entries of $\mathbf{b}$. 
	It is clear that $\sum_{i=1}^n b_i = 0$. On the other hand, 
	$$
	\prod_{j \in [n], j \neq i}(b_j - b_i) = 
	\prod_{j \in [n], j \neq i}(a_j - a_i), \, \forall i \in [n].
	$$
	The result follows from \autoref{theorem:GRS_self_dual} and 
	\autoref{theorem:self_dual_1}.
\end{proof}
Similarly, we have the following characterization of when $\GRS_{k,k-1}(\mathbf{a},\mathbf{v})$ is self-dual. 
\begin{theorem}
\label{theorem:self_dual_2}
Let $n = 2k \ge 8$, and $\mathbf{a} = (a_1,\dots,a_n)$ be pairwise distinct 
evaluation points in $\F_q$. Then, there exist non-zero factors $\mathbf{v} 
= (v_1,\dots,v_n) \in (\F_q^*)^n$ such that 
$\GRS_{k,k-1}(\mathbf{a},\mathbf{v})$ 
is self-dual if and only if $t_{n-1} = 0$ and 
$u_1s_{n-2}(\mathbf{a}_1),\dots,u_ns_{n-2}(\mathbf{a}_n)$ 
are simultaneously squares or non-squares in $\F_q$, where 
$\mathbf{a}_i \defeq \mathbf{a} \setminus \setnd{a_i}$.
\end{theorem}
\begin{proof}
	$\implies$: Suppose $\GRS_{k,k-1}(\mathbf{a},\mathbf{v})$ is self-dual for 
	some $\mathbf{v} = (v_1,\dots,v_n) \in 
	(\F_q^*)^n$. Let $G$ be the generator matrix for 
	$\GRS_{k,k-1}(\mathbf{a},\mathbf{v})$ as in 
	\autoref{eqn:matrix_subcode_GRS}, and 
	\begin{equation}
%		\label{eqn:k_1_powers}
	M = \spalignmat{
		a_1^0,\dots,a_n^0;
		a_1^2,\dots,a_n^2;
		\vdots,\vdots,\vdots;
		a_1^{2k},\dots,a_n^{2k}
	}.
	\end{equation}
	We have that $GG^\intercal = 0$, which is equivalent to 
	that $M\mathbf{v}_2 = 0$, where $\mathbf{v}_2 = (v_1^2,\dots,v_n^2)$. 
	As $\mathbf{v}_2$ is non-zero and is in the kernel of $M$, $M$ is not 
	invertible. Thus, there exist 
	$f_0,f_2,\dots,f_{2k} \in \F_q$ not all zero, such that 
	$(f_0,f_2,\dots,f_{2k})M = 0$. Let $f(x) = 
	f_0+f_2x^2+\dots+f_{2k}x^{2k}$, then $a_1,\dots,a_n$ are zeros 
	of $f(x)$. 
	So the degree of $f(x)$ is forced to be $2k$, i.e., $f_{2k} \neq 0$. By a 
	scaling, we may assume that 
	$f_{2k} = 1$. Then, $f(x) = f_{\mathbf{a}}(x) \defeq \prod_{i=1}^n(x-a_i)$.
	Comparing the coefficients of $x$, we have that 
	$t_{n-1} = s_{n,n-1}(\mathbf{a}) = 0$.
	
	From the argument above, we also know that $M$ is of rank $2k-1$, as any 
	non-zero solution $(f_0,f_2,\dots,f_{2k})$ of 
	the linear equation $\mathbf{y}M = 0$ is a multiple of 
	$(f_\mathbf{a}[0],f_{\mathbf{a}}[2],\dots,f_{\mathbf{a}}[2k])$, where 
	$f_\mathbf{a}[i]$ is the coefficient of $x^i$ in $f_{\mathbf{a}}(x)$. 
	Therefore, $\mathbf{v}_2$ spans the kernel of $M$. By \autoref{lemma:sum2}, 
	$\mathbf{w} = (u_1s_{n-2}(\mathbf{a}_1),\dots,u_ns_{n-2}(\mathbf{a}_n))$ 
	also lives in the kernel of $M$, and is non-zero as $\sum_{i=1}^na_iu_1s_{n-2}(\mathbf{a}_i) = 
	1$ by \autoref{lemma:sum2}. So, there is $\lambda \in \F_q^*$ such that $\mathbf{w} = \lambda 
	\mathbf{v}_2$. We conclude that $u_1s_{n-2}(\mathbf{a}_1),\dots,u_ns_{n-2}(\mathbf{a}_n)$ 
	are simultaneously squares or non-squares in $\F_q$.
	
	$\impliedby$: The proof of the converse implication is similar to that of 
	\autoref{theorem:self_dual_1}. 
	Let $w_i = u_is_{n-2}(\mathbf{a}_i)$, and assume 
	that $w_1,\dots,w_n$ are 
	all squares or non-squares. Take $\lambda \in \F_q^*$ such that $\lambda 
	w_1,\dots,\lambda w_n$ are all squares. 
	Then, $\GRS_{k,k-1}(\mathbf{a},(\sqrt{\lambda w_i})_{i \in [n]})$ is 
	self-dual by \autoref{lemma:sum2}.
\end{proof}
The following lemma further ensures that the equivalent condition in \autoref{theorem:self_dual_2}
for $\GRS_{k,k-1}(\mathbf{a},\mathbf{v})$ being self-dual is well-defined.
\begin{lemma}
	\label{lemma:condition_well_defined}
	Let $\mathbf{a} = (a_1,\dots,a_n)$ be pairwise distinct evaluation points in $\F_q$. 
	If $t_{n-1} \defeq s_{n-1}(\mathbf{a}) = 0$, then $0 \notin \mathbf{a}$, and 
	$s_{n-2}(\mathbf{a}_1),\dots,s_{n-2}(\mathbf{a}_n)$ are all non-zero, where 
	$\mathbf{a}_i = \mathbf{a} \setminus \setnd{a_i}$.
\end{lemma}
\begin{proof}
	If $0 \in \mathbf{a}$, then 
	$t_{n-1} = s_{n-1}(\mathbf{a}) = s_{n-1}(\mathbf{a} \setminus \setnd{0}) \neq 0$, contradicting 
	to the assumption that $t_{n-1} = 0$. On the other hand, note that 
	$$
	0 = t_{n-1} = s_{n-1}(\mathbf{a}) = a_is_{n-2}(\mathbf{a}_i) + s_{n-1}(\mathbf{a}_i).
	$$
	Thus, $a_is_{n-2}(\mathbf{a}_i) = -s_{n-1}(\mathbf{a}_i)$. As $0 \notin \mathbf{a}$, we have that 
	$a_i \neq 0$, $s_{n-1}(\mathbf{a}_i) \neq 0$, and we conclude that $s_{n-2}(\mathbf{a}_i) \neq 0$. 
\end{proof}
%It was already shown in \cite{subcodeEGRS} that 
%$\GRS_{k,1}(\mathbf{a},\mathbf{v},\infty)$ can never be self-dual for any 
%evaluation points $\mathbf{a}$ and factors $\mathbf{v}$, when $n+1 = 2k$. 
Next, we characterize when $\GRS_{k,k-1}(\mathbf{a},\mathbf{v},\infty)$ is 
self-dual.
\begin{theorem}
\label{theorem:self_dual_3}
Let $n + 1 = 2k \ge 8$, and $\mathbf{a} = (a_1,\dots,a_n)$ be pairwise distinct 
evaluation points in $\F_q$. Then, there exist non-zero factors $\mathbf{v} 
= (v_1,\dots,v_n) \in (\F_q^*)^n$ such that 
$\GRS_{k,k-1}(\mathbf{a},\mathbf{v},\infty)$ 
is self-dual if and only if $t_{n-1} = 0$ and 
$t_nu_1s_{n-2}(\mathbf{a}_1),\dots,t_nu_ns_{n-2}(\mathbf{a}_n)$ 
are all squares in $\F_q$, where $\mathbf{a}_i \defeq 
\mathbf{a} \setminus \setnd{a_i}$.
\end{theorem}
\begin{proof}
The proof is similar to that of \autoref{theorem:self_dual_2}. 

$\implies$: Suppose $\GRS_{k,k-1}(\mathbf{a},\mathbf{v},\infty)$ 
is self-dual for some $\mathbf{v} = (v_1,\dots,v_n) \in 
(\F_q^*)^n$. Let $G$ be the generator matrix for 
$\GRS_{k,k-1}(\mathbf{a},\mathbf{v},\infty)$ as in 
\autoref{eqn:matrix_subcode_EGRS}, and 
\begin{equation}
%	\label{eqn:k_1_powers}
	M = \spalignmat{
		a_1^0,\dots,a_n^0;
		a_1^2,\dots,a_n^2;
		\vdots,\vdots,\vdots;
		a_1^{2k-1},\dots,a_n^{2k-1}
	}.
\end{equation}
We have that $GG^\intercal = 0$, which is equivalent to that
$M\mathbf{v}_2 = 0$ and $\sum_{i=1}^na_i^{2k}v_i^2 + 1 = 0$, where 
$\mathbf{v}_2 = (v_1^2,\dots,v_n^2)$. 
By a similar argument as in the proof of \autoref{theorem:self_dual_2}, 
we know that $t_{n-1} = 0$, $M$ is of rank $n-1$ and 
$\mathbf{w} = \lambda \mathbf{v}_2$ for some $\lambda \in \F_q^*$, 
where $\mathbf{w} = 
(u_1s_{n-2}(\mathbf{a}_1),\dots,u_ns_{n-2}(\mathbf{a}_n))$. 
By \autoref{lemma:sum2}
$$
\sum_{i=1}^na_i^{2k}v_i^2 + 1 = 
\frac{1}{\lambda} \sum_{i=1}^na_i^{2k}u_is_{n-2}(\mathbf{a}_i) + 1 = 
\frac{1}{\lambda}(t_1t_{n-1}-t_n) + 1 = 
1 - \frac{t_n}{\lambda} = 0,
$$
and $\lambda = t_n$. Therefore, $t_nu_1s_{n-2}(\mathbf{a}_1),\dots,t_nu_ns_{n-2}(\mathbf{a}_n)$ are 
all squares. 

$\impliedby$: Assume $t_{n-1} = 0$, by \autoref{lemma:condition_well_defined}, $t_n \neq 0$. Let 
$$\mathbf{v} \defeq (v_1,\dots,v_n) \defeq 
\frac{1}{t_n}(u_1s_{n-1,n-2}(\mathbf{a}_1),\dots,u_ns_{n-1,n-2}(\mathbf{a}_n)).$$
 
By assumption, $v_1,\dots,v_n$ are all sqaures. Then, by 
\autoref{lemma:sum2}, 
$\GRS_{k,k-1}(\mathbf{a},(\sqrt{v_i})_{i \in [n]},\infty)$ is self-dual.
\end{proof}
We finish this section with an example of self-dual $\GRS_{k,k-1}(\mathbf{a},\mathbf{v})$. 
We will show in \autoref{corollary:subcode_GRS_always_NMDS} that 
$\GRS_{k,k-1}(\mathbf{a},\mathbf{v})$ is always NMDS or MDS, and the resuting code is 
self-dual NMDS or self-dual MDS.
\begin{example}
Let $q \equiv 1 \pmod{4}$ and $n \mid (q-1)$ be even. Let $\mathbf{a} = (a_1,\dots,a_n)$ be the 
elements of the cyclic subgroup of $\F_q^*$ of order $n$. Then $f_{\mathbf{a}}(x) \defeq 
\prod_{i=1}^n(x-a_i) = x^n-1$, and $\frac{1}{u_i} = \prod_{j \in [n], j \neq i}(a_i-a_j) = f_{\mathbf{a}}^\prime(a_i) = 
na_i^{n-1}$. Let $\mathbf{a}_i = \mathbf{a} \setminus \setnd{a_i}$ and 
$f_{\mathbf{a}_i}(x) \defeq \prod_{j \in [n], j \neq i}(x-a_j) = \frac{f_{\mathbf{a}}(x)}{(x-a_i)}$. 
Note that 
$$
f_{\mathbf{a}}(x) = x^n - t_1x^{n-1} + \dots + (-1)^{n-1}t_{n-1}x + (-1)^nt_n = x^n-1,
$$
and thus $t_1 = 0$. Similarly, 
$$
f_{\mathbf{a}_i}(x) = x^{s-1} - s_{1}(\mathbf{a}_i)x^{n-2} + \dots + (-1)^{n-2}s_{n-2}(\mathbf{a}_i)x + 
(-1)^{n-1}s_{n-1}(\mathbf{a}_i), 
$$
and $s_{n-2}(\mathbf{a}_i) = (-1)^{n-2}f_{\mathbf{a}_i}^\prime(0)$. Note that 
$f_{\mathbf{a}_i}^\prime(x) = \frac{f_{\mathbf{a}}^\prime(x)(x-a_i) - f_{\mathbf{a}_i}(x)}{(x-a_i)^2}$. 
Therefore, 
$$
u_is_{n-2}(\mathbf{a}_i) = \frac{1}{na_i^{n+1}} = \frac{1}{na_i}.
$$
As $a_1,\dots,a_n$ are all squares in $\F_q$, $u_is_{n-2}(\mathbf{a}_i), i \in [n]$ are all squares 
(when $n$ is a square in $\F_q$), or all non-squares (when $n$ is a non-square in $\F_q$). By 
\autoref{theorem:self_dual_2}, there are non-zero factors $\mathbf{v} \in (\F_q^*)^n$ such that 
$\GRS_{n/2,n/2-1}(\mathbf{a},\mathbf{v})$ is self-dual.
\end{example}

\section{NMDS Subcodes of GRS Codes}
\label{sec:NMDS}
In this section, we determine when $\GRS_{k,r}(\mathbf{a},\mathbf{v})$ or 
$\GRS_{k,r}(\mathbf{a},\mathbf{v},\infty)$ is NMDS. 
Let $G$ be a generator matrix of an arbitrary $[n,k]_q$ code $C$, and 
$\mathbf{w} \in \dual{C}$ be non-zero. Then, $G\mathbf{w}^\intercal = 0$, 
and the columns in $G$ at the support of $\mathbf{w}$ are linearly 
dependent. In particular, $\dual{C}$ has minimum distance at least 
$\dual{d}$ if and only if any $\dual{d}-1$ columns of $G$ are linearly independent. 
This characterization of the minimum distance of the dual codes will 
be crucial to us. 

It is clear that the NMDS property of 
$\GRS_{k,r}(\mathbf{a},\mathbf{v})$ or $\GRS_{k,r}(\mathbf{a},\mathbf{v},\infty)$ 
is independent of the choice of $\mathbf{v} \in (\F_q^*)^n$. So, in this section, 
we will always assume that $\mathbf{v} = (1,\dots,1)$, and write 
$\GRS_{k,r}(\mathbf{a}) \defeq \GRS_{k,r}(\mathbf{a},(1,\dots,1))$, and 
$\GRS_{k,r}(\mathbf{a},\infty) \defeq \GRS_{k,r}(\mathbf{a},(1,\dots,1),\infty)$. 

For $r = 1$, it was shown in \cite{Han2024NMDS} and \cite{subcodeEGRS} 
respectively, that $\GRS_{k,1}(\mathbf{a})$ 
and $\GRS_{k,1}(\mathbf{a},\infty)$ are always NMDS or 
MDS. So, in the following, we will focus on the case $r \in [2,k-1]$. 
First, we characterize when $\GRS_{k,r}(\mathbf{a})$ is NMDS. 
To this end, we consider the rank of any $k-1$ columns of the generator 
matrix in \autoref{eqn:matrix_subcode_GRS} of 
$\GRS_{k,r}(\mathbf{a})$. 
\begin{lemma}
	\label{lemma:nmds}
Let $a_1,\dots,a_{k-1} \in \F_q$ be pairwise distinct. For $r \in [2,k-1]$, let 
\begin{equation}
	\label{eqn:Gr}
G_{r} = 
\spalignmat{
a_1^0,\dots,a_{k-1}^0;
\vdots,\vdots,\vdots;
a_1^{k-r-1},\dots,a_{k-1}^{k-r-1};
a_1^{k-r+1},\dots,a_{k-1}^{k-r+1};
\vdots,\vdots,\vdots;
a_1^{k-1},\dots,a_{k-1}^{k-1};
a_1^k,\dots,a_{k-1}^k
}_{k \times (k-1)}.
\end{equation}
Then 
\begin{enumerate}[label=(\arabic*)]
\item $G_{r}$ has rank at least $k-2$, \label{item:dualdist1}
\item $G_{r}$ has rank $k-1$ if and only if $s_{k-1,r-1}(a_1,\dots,a_{k-1})$ 
and $s_{k-1,r}(a_1,\dots,a_{k-1})$ are not both zero.
\label{item:dualdist2}
\end{enumerate}
\end{lemma}
\begin{proof}
\ref{item:dualdist1}: The first $k-2$ rows of $G_r$ constitute 
a submatrix of the invertible Vandermonde matrix $V(a_1,\dots,a_{k-1})$ 
defined in \autoref{eqn:Vandermonde}, and thus are linearly independent. 
So $G_r$ has rank at least $k-2$.

\ref{item:dualdist2}:  $\phi$ be the following linear isomorphism
$$
\phi : \F_q[x]_{k,r} \to \F_q^k, \, f(x) \mapsto (f[0],\dots,f[k-r-1],f[k-r+1],\dots,f[k]),
$$
and $\psi$ be the following linear map 
$$
\psi : \F_q^k \to \F_q^{k-1}, \, \mathbf{w} \mapsto 
\mathbf{w}G_r.
$$
Consider the composition 
$$
\psi \circ \phi : \F_q[x]_{k,r} \to \F_q^k \to \F_q^{k-1}.
$$
Note that the composition $\psi \circ \phi$ coincides with the 
evaluation map, i.e., $\psi \circ \phi(f(x)) = (f(a_1),\dots,f(a_{k-1}))$ 
for all $f(x) \in \F_q[x]_{k,r}$. It is also clear that 
\begin{center}
\begin{tabular}{lll}
$G_r$ has rank $k-1$ & $\iff$ & $\psi \circ \phi$ is surjective \\
& $\iff$ & the kernel of $\psi \circ \phi$ is of dimension $1$ \\
& $\iff$ & there is a unique monic polynomial in $\F_q[x]_{k,r}$ vanishing on 
$a_1,\dots,a_{k-1}$.
\end{tabular}
\end{center}
Let $g(x) = \prod_{i=1}^{k-1}(x-a_i)$ and $h_b(x) = g(x)(x-b), b \in \F_q$. 
Then, $g(x)$ and $h_b(x), b \in \F_q$ are all the monic polynomials of degree at most $k$ 
vanishing on $a_1,\dots,a_{k-1}$, which possibly live in $\F_q[x]_{k,r}$. Note that 
\begin{align*}
g(x) \in \F_q[x]_{k,r} &\iff g[k-r] = (-1)^{r-1}s_{k-1,r-1}(a_1,\dots,a_{k-1}) = 0, \\
h_b(x) \in \F_q[x]_{k,r} &\iff h_b[k-r] = (-1)^{r}s_{k,r}(a_1,\dots,a_{k-1},b) = 0.
\end{align*}
We seek when exactly one of $g(x)$ and $h_b(x), b \in \F_q$ lives in $\F_q[x]_{k,r}$, i.e., 
when exactly one of $g[k-r]$ and $h_b[k-r], b \in \F_q$ equals zero. Note that 
$$
s_{k,r}(a_1,\dots,a_{k-1},b) = 
s_{k-1,r}(a_1,\dots,a_{k-1}) + bs_{k-1,r-1}(a_1,\dots,a_{k-1}).
$$
If $s_{k-1,r-1}(a_1,\dots,a_{k-1}) \neq 0$, there is a unique 
$b_0 \in \F_q$ such that $h_{b_0}[k-r] = 0$, and then exactly one of $g[k-r]$ and $h_b[k-r], b \in \F_q$ is 
zero. If $s_{k-1,r-1}(a_1,\dots,a_{k-1}) = 0$, we must have $s_{k-1,r}(a_1,\dots,a_{k-1}) \neq 0$, and then 
exactly one of $g[k-r]$ and $h_b[k-r], b \in \F_q$ is zero.
\end{proof}
\autoref{lemma:nmds} immediately implies the following 
characterization of when $\GRS_{k,r}(\mathbf{a})$ is NMDS, for $r \in [2,k-1]$.
\begin{theorem}
	\label{theorem:nmds}
Let $\mathbf{a}=(a_1,\dots,a_n)$ be pairwise distinct evaluation points in $\F_q$. For $r \in [2,k-1]$, 
let $\dual{d}$ be the minimum distance of the dual code of $\GRS_{k,r}(\mathbf{a})$. 
Then, $\dual{d} \ge k$ if and only if for any subset $A \subset \mathbf{a}$ of size $k-1$, $s_{k-1,r-1}(A)$ and 
$s_{k-1,r}(A)$ are not both zero, in which case $\GRS_{k,r}(\mathbf{a})$ is NMDS or MDS. 
\end{theorem}
\begin{proof}
$\dual{d} \ge k$ if and only if any $k-1$ columns of $G_{k,r}(\mathbf{a})$ have rank $k-1$, where 
$G_{k,r}(\mathbf{a})$ is the generator matrix of $\GRS_{k,r}(\mathbf{a})$ defined in \autoref{eqn:matrix_subcode_GRS}. 
As any $k-1$ columns of $G_{k,r}(\mathbf{a})$ is of the form in \autoref{eqn:Gr}, the 
result follows immediately from \autoref{lemma:nmds}. 
\end{proof}
As an immediate consequence of \autoref{theorem:nmds}, we know that $\GRS_{k,k-1}(\mathbf{a})$ is 
always NMDS or NMDS.
\begin{corollary}
	\label{corollary:subcode_GRS_always_NMDS}
Let $\mathbf{a}=(a_1,\dots,a_n)$ be pairwise distinct evaluation points in $\F_q$. Then 
$\GRS_{k,k-1}(\mathbf{a})$ is always NMDS or MDS.
\end{corollary}
\begin{proof}
Let $A \subset \mathbf{a}$ be of size $k-1$. If $0 \in A$, then $s_{k-1,k-2}(A) \neq 0$. 
On the other hand, if $0 \notin A$, then $s_{k-1,k-1}(A) \neq 0$. The result follows from 
\autoref{theorem:nmds}.
\end{proof}
Next, we characterize when $\GRS_{k,r}(\mathbf{a},\infty)$ is NMDS, for $2 \in [2,k-1]$. 
We split into two cases where $r = 2$ or $r \in [3,k-1]$.
\begin{theorem}
Let $\mathbf{a} = (a_1,\dots,a_n)$ be pairwise distinct evaluation points in $\F_q$, and 
$\dual{d}$ be the minimum distance of the dual code of $\GRS_{k,2}(\mathbf{a},\infty)$. 
Then, $\dual{d} \ge k$ if and only if for any subset $A \subset \mathbf{a}$ of size $k-1$, $s_{k-1,1}(A)$ and 
$s_{k-1,2}(A)$ are not both zero, in which case $\GRS_{k,2}(\mathbf{a},\infty)$ is NMDS or MDS. 
\end{theorem}
\begin{proof}
	Let $G_{k,2}(\mathbf{a},\infty)$ be the generator matrix of $\GRS_{k,2}(\mathbf{a},\infty)$ defined in 
	\autoref{eqn:matrix_subcode_EGRS}, and $N$ be the submatrix of $G_{k,2}(\mathbf{a},\infty)$ consisting 
	of any $k-1$ columns. If $N$ is chose from the first $n$ columns of $G_{k,2}(\mathbf{a},\infty)$, then 
	$N$ is of the form in \autoref{eqn:Gr} and \autoref{lemma:nmds} applies. 
	If the last column of $G_{k,2}(\mathbf{a},\infty)$ is chosen in $N$, then without loss of generality, $N$ is 
	of the form 
	$$
	N = 
	\spalignmat[c]{
		a_1^0,\dots,a_{k-2}^0,0;
		\vdots,\vdots,\vdots,\vdots;
		a_1^{k-3},\dots,a_{k-2}^{k-3},0;
		a_1^{k-1},\dots,a_{k-2}^{k-1},0;
		a_1^{k},\dots,a_{k-2}^{k},1;
	}.
	$$
	The first $(k-2)$ rows and $(k-2)$ columns of $N$ constitute the invertible Vandermonde matrix 
	$V(a_1,\dots,a_{k-2})$. Also, the last row of $N$ is not in the span of the first $(k-2)$ rows of $N$. 
	We conclude that $N$ has rank $k-1$.
\end{proof}

\begin{lemma}
	\label{lemma:NMDS_EGRS}
Let $a_1,\dots,a_{k-2} \in \F_q$ be pairwise distinct. For $r \in [3,k-1]$, let 
\begin{equation}
	\label{eqn:Gr_EGRS}
	N_r = \spalignmat[c]{
		a_1^0,\dots,a_{k-2}^0,0;
		\vdots,\vdots,\vdots;
		a_1^{k-r-1},\dots,a_{k-2}^{k-r-1},0;
		a_1^{k-r+1},\dots,a_{k-2}^{k-r+1},0;
		\vdots,\vdots,\vdots;
		a_1^{k-2},\dots,a_{k-2}^{k-2},0;
		a_1^{k-1},\dots,a_{k-2}^{k-1},0;
		a_1^k,\dots,a_{k-2}^k,1
	}_{k \times (k-1)}.
\end{equation}
Then 
\begin{enumerate}[label=(\arabic*)]
\item $N_r$ has rank at least $k-2$,
\item $N_r$ has rank $k-1$ if and only if $s_{k-2,r-2}(a_1,\dots,a_{k-2})$ and $s_{k-2,r-1}(a_1,\dots,a_{k-2})$ are 
not both zero. 
\end{enumerate}
\end{lemma}
\begin{proof}
Note that the submatrix of $N_r$ consisting of the first $(k-1)$ rows and $(k-2)$ columns, is 
of the form in \autoref{eqn:Gr}, but with a smaller size. And the result follows from \autoref{lemma:nmds}.
\end{proof}
\begin{theorem}
Let $\mathbf{a} = (a_1,\dots,a_n)$ be pairwise distinct evaluation points in $\F_q$. 
For $r \in [3,k-1]$, let $\dual{d}$ be the minimum distance of the dual code of  
$\GRS_{k,r}(\mathbf{a},\infty)$. Then, $\dual{d} \ge k$ if and only if for any subset 
$A \subset \mathbf{a}$ of size $k-1$, $s_{k-1,r-1}(A)$ and $s_{k-1,r}(A)$ are not both zero, 
and for any subset $B \subset \mathbf{a}$ of size $k-2$, $s_{k-2,r-2}(B)$ and $s_{k-2,r-1}(B)$ are not 
both zero, in which case $\GRS_{k,r}(\mathbf{a},\infty)$ is NMDS or MDS.
\end{theorem}
\begin{proof}
Let $N$ be any $k-1$ columns of the generator matrix $G_{k,r}(\mathbf{a},\infty)$ defined in 
\autoref{eqn:matrix_subcode_EGRS}. If $N$ is chosen from the first $n$ columns of $G_{k,r}(\mathbf{a},\infty)$, 
then $N$ is of the form in \autoref{eqn:Gr}, and \autoref{lemma:nmds} applies. 
If the last column of $G_{k,r}(\mathbf{a},\infty)$ is chosen in $N$, then $N$ is of the form in 
\autoref{eqn:Gr_EGRS}, and \autoref{lemma:NMDS_EGRS} applies.
\end{proof}
Similar to \autoref{corollary:subcode_GRS_always_NMDS}, $\GRS_{k,k-1}(\mathbf{a},\infty)$ is also 
always NMDS or MDS.
\begin{corollary}
	\label{corollary:subcode_EGRS_alwasy_NMDS}
Let $\mathbf{a} = (a_1,\dots,a_n)$ be pairwise distinct evaluation points in $\F_q$. Then 
$\GRS_{k,k-1}(\mathbf{a},\infty)$ is always NMDS or MDS.
\end{corollary}
The proof of \autoref{corollary:subcode_EGRS_alwasy_NMDS} is similar to that of 
\autoref{corollary:subcode_GRS_always_NMDS}, and is omitted.

\section{Dual Codes of the Subcodes of GRS Codes}
\label{sec:dual_code_subcodes}
In this section, we find out the dual codes of $\GRS_{k,r}(\mathbf{a},\mathbf{v})$ and 
$\GRS_{k,r}(\mathbf{a},\mathbf{v},\infty)$ for $r \in \setnd{1,k-1}$ or $r = 2$. 
Let $G$ be a generator matrix of any $[n,k]_q$ code and $H$ be a parity 
check matrix of $G$. For any invertible diagonal $n \times n$ matrix $D$, 
the code with generator matrix $GD$ has parity check matrix 
$HD^{-1}$. So, it suffices to only consider the case $\mathbf{v} = 
(1,\dots,1)$. As in the previous section, we write 
$\GRS_{k,r}(\mathbf{a}) \defeq \GRS_{k,r}(\mathbf{a},(1,\dots,1))$ and 
$\GRS_{k,r}(\mathbf{a},\infty) \defeq 
\GRS_{k,r}(\mathbf{a},(1,\dots,1),\infty)$. 

\subsection{The case $r \in \setnd{1,k-1}$}
For $r = 1$, the parity check matrix of $\GRS_{k,1}(\mathbf{a},\infty)$ 
was already determined in \cite{subcodeEGRS} (see \autoref{eqn:subcode1_EGRS_dual}). 
If $t_1 \neq 0$, the dual code of 
$\GRS_{k,1}(\mathbf{a})$ is a $(+)$-TGRS code \cite{Huang2021twistedRS} 
(see \autoref{eqn:subcode1_GRS_dual_t1_nonzero}). 
So, we only focus on the dual code of $\GRS_{k,1}(\mathbf{a})$ with $t_1 = 0$.
\begin{theorem}
	\label{theorem:dual_code_1}
Let $2 \le k \le n-2$ and $\mathbf{a} = (a_1,\dots,a_n)$ be 
$n$ pairwise distinct evaluation points in $\F_q$. If $t_1 = 0$, 
the dual code of $\GRS_{k,1}(\mathbf{a})$ equals 	
$\GRS_{n-k,1}(\mathbf{a},(u_i)_{i \in [n]})$.
\end{theorem}
\begin{proof}
	Let $\mathbf{c}_l \defeq (a_i^l)_{i \in [n]}$, $l \in J(k,1)$, be the rows of the 
	generator matrix of $\GRS_{k,1}(\mathbf{a})$ as in \autoref{eqn:matrix_subcode_GRS}, and 
	$\mathbf{b}_m \defeq (u_ia_i^m)_{i \in [n]}, m \in J(n-k,1)$ be the rows of the generator matrix 
	of $\GRS_{n-k,1}(\mathbf{a},(u_i)_{i \in [n]})$. By calculation, 
%	\begin{equation}
	$\mathbf{c}_l \cdot \mathbf{b}_m = \sum_{i=1}^n u_ia_i^{l+m}$.
%	\end{equation}
	As $J(k,1) + J(n-k,1) \subset [0,n]$ and $(n-1) \notin J(k,1) + J(n-k,1)$, 
	we know that $\mathbf{c}_l \cdot \mathbf{b}_m = 0$ for all $l 
	\in J(k,1), m \in J(n-k,1)$, by \autoref{lemma:sum} and 
	the assumption that $t_1 = 0$. On the other hand, 
	$\GRS_{k,1}(\mathbf{a})$ is of dimension $k$ and 
	$\GRS_{n-k,1}(\mathbf{a},(u_i)_{i \in [n]})$ is of dimension 
	$n-k$. We conclude that the dual code of 
	$\GRS_{k,1}(\mathbf{a},\mathbf{v})$ is 
	$\GRS_{n-k,1}(\mathbf{a},(u_i)_{i \in [n]})$.
\end{proof}
By a similar argument as in the proof of \autoref{theorem:dual_code_1} and 
\autoref{lemma:sum2}, we can determine the dual code of $\GRS_{k,k-1}(\mathbf{a})$ and 
$\GRS_{k,k-1}(\mathbf{a},\infty)$ when $t_{n-1} = 0$. 
\begin{theorem}
	\label{theorem:dual_code_k_1}
Let $3 \le k \le n-2$ and $\mathbf{a} = (a_1,\dots,a_n)$ be $n$ pairwise distinct evaluation points in $\F_q$. 
Assume that $t_{n-1} = 0$. Then, the dual code of $\GRS_{k,k-1}(\mathbf{a})$ is 
$\GRS_{n-k,n-k-1}(\mathbf{a},(u_is_{n-2}(\mathbf{a}_i))_{i \in [n]})$, where 
$\mathbf{a}_i \defeq \mathbf{a} \setminus \setnd{a_i}$, and $\GRS_{k,k-1}(\mathbf{a},\infty)$ has the 
following parity check matrix
\begin{equation}
	\label{eqn:parity_check_SEGRS_k_1}
\spalignmat{
	a_1^0,\dots,a_n^0,0;
	a_1^2,\dots,a_n^2,0;
	\vdots,\vdots,\vdots,\vdots;
	a_1^{n-k},\dots,a_n^{n-k},0;
	a_1^{n-k+1},\dots,a_n^{n-k+1},t_n
}D(u_1s_{n-2}(\mathbf{a}_1),\dots,u_ns_{n-2}(\mathbf{a}_n),1).	
\end{equation}
%where $D(u_1s_{n-2}(\mathbf{a}_1),\dots,u_ns_{n-2}(\mathbf{a}_n),1)$ is the $(n+1) \times (n+1)$ diagonal 
%matrix with entries $u_1s_{n-2}(\mathbf{a}_1),\dots,u_ns_{n-2}(\mathbf{a}_n)$ and $1$.
\end{theorem}
\begin{proof}
We note that $s_{n-2}(\mathbf{a}_i), i \in [n]$ are all non-zero by \autoref{lemma:condition_well_defined} and 
the assumption that $t_{n-1} = 0$. Therefore, 
$\GRS_{n-k,n-k-1}(\mathbf{a},(u_is_{n-2}(\mathbf{a}_i))_{i \in [n]})$ and the parity check matrix 
in \autoref{eqn:parity_check_SEGRS_k_1} are well-defined. With \autoref{lemma:sum2}, \autoref{theorem:dual_code_k_1} 
can be checked in a similar way as in the proof of \autoref{theorem:dual_code_1}. The details are omitted.
\end{proof}

\subsection{The case $r = 2$}
In this section, we find out the dual codes of 
$\GRS_{k,2}(\mathbf{a})$ and $\GRS_{k,2}(\mathbf{a},\infty)$.
Recall that $h_i \defeq \sum_{i=1}^n u_ia_i^{n-1+i}$ for $i \ge 0$, and 
$h_0 = 1$, $h_1 = t_1$, $h_2 = t_1^2-t_2$ (see \autoref{lemma:sum}). 
%By \autoref{lemma:sum},  and 
%$h_l$ is a polynomial in $t_i, i \in [n]$ for $l \ge 2$.
\begin{theorem}
	\label{theorem:dual_code_2}
Let $3 \le k \le n-2$, and $\mathbf{a} = (a_1,\dots,a_n)$ be $n$ pairwise 
distinct evaluation points in $\F_q$. Let $\mathbf{u} = (u_1,\dots,u_n)$. 
Then, the dual code of $\GRS_{k,2}(\mathbf{a})$ equals
$$
\evmap_{\mathbf{a},\mathbf{u}}\qty(
\spn\setnd{1,x,x^2,\dots,x^{n-k-2},g(x)})
$$
where 
	\begin{enumerate}[label=(\arabic*)]
		\item $g(x) = x^{n-k+1}$, if $h_1 = h_2 = 0$; \label{item:case1}
		\item $g(x) = x^{n-k-1}-x^{n-k+1}/h_2$, if $h_1=0$ and $h_2 \neq 0$; 
		\label{item:case2}
		\item $g(x) = x^{n-k-1}-x^{n-k}/h_1 + x^{n-k+1}/h_1^2$, if $h_1 \neq 0$ 
		and 
		$h_2 = 0$; 
		\label{item:case3}
		\item $g(x) = x^{n-k}-x^{n-k+1}/h_1$, if $h_1 \neq 0$, $h_2 \neq 0$ and 
		$h_2 
		= h_1^2$; \label{item:case4}
		\item $g(x) = x^{n-k-1} - \frac{h_1x^{n-k}-x^{n-k+1}}{h_1^2-h_2}$, if 
		$h_1 \neq 0$, $h_2 \neq 0$ and $h_2 \neq h_1^2$. \label{item:case5}
	\end{enumerate}
\end{theorem}
We note that in case \ref{item:case1} of \autoref{theorem:dual_code_2}, 
the dual code is again a subcode of the GRS 
code, but it is not of the form (at least superficially)  
$\GRS_{k,r}(\mathbf{a},\mathbf{v})$ in the scope of this paper. 
In cases \ref{item:case2} \ref{item:case3} and \ref{item:case5}, the dual 
code is a TGRS code. Note that the dual code in \ref{item:case4} is not 
(at least superficially) a TGRS code. 
\begin{proof}[Proof of \autoref{theorem:dual_code_2}]
	The proof follows a similar approach adopted in \cite{Huang2021twistedRS}. 
	Let 
	$$A = \spalignmat[c]{
		a_1^0,a_2^0,\dots,a_n^0 ;
		a_1^1,a_2^1,\dots,a_n^1 ;
		\vdots,\vdots,\ddots,\vdots;
		a_1^{k},a_2^{k},\dots,a_n^{k}
	}_{(k+1)\times n}
	\quad 
	B = \spalignmat[c]{
		u_1a_1^{n-1}, u_1a_1^{n-2},\dots,u_1a_1^0;
		u_2a_2^{n-1}, u_2a_2^{n-2},\dots,u_2a_1^0;
		\vdots,\vdots,\ddots,\vdots;
		u_na_n^{n-1}, u_na_n^{n-2},\dots,u_na_n^0
	}_{n \times n}
	$$ 
	By \autoref{lemma:sum}, 
	$
	AB = \spalignmat{C_{(k+1)\times (k+1)},0_{(k+1)\times (n-k-1)}},
	$
	where 
	$$
	C = 
	\spalignmat{
		1;
		\vdots,\ddots;
		h_{k-2},\dots,1;
		h_{k-1},\dots,h_1,1;
		h_{k},\dots,h_2,h_1,1
	}_{(k+1) \times (k+1)}.
	$$
	Note that the rows and columns of $A,B,C$ are indexed from $0$. We will 
	perform an invertible linear transformation 
	$Q_{(k+1) \times (k+1)}$ on the columns of $C$, so that the last column of 
	$C$ became all zero except for the $(k-2)$-th entry in the last column of 
	$C$. The linear transformation $Q$ reflected on $B$ is represented by 
	$$
	P = \spalignmat{
		Q, ; , I}_{n \times n},
	$$
	i.e., $ABP = \spalignmat{CQ,0}$. Then the last $n-k$ columns of $BP$ will 
	span the dual code of $\GRS_{k,2}(\mathbf{a})$.
	
	\textbf{Case} \ref{item:case1}: 
	In this case, we just swap the $(k-2)$-th and the $k$-th columns of $C$, 
	which corresponds to swapping the $(k-2)$-th and $k$-th columns of $B$.

	\textbf{Case} \ref{item:case2}: In this case, we 
	add $(-\frac{1}{h_2})$-times of the $(k-2)$-th column of $C$ to 
	the $k$-th column. 
	
	\textbf{Case} \ref{item:case3}: In this case, we 
	first add $(-\frac{1}{h_1})$-times of the $(k-2)$-th of $C$ to 
	the $(k-1)$-th column, and then add $(-\frac{1}{h_1})$-times of 
	the $(k-2)$-th column to the $k$-th column. Focusing on 
	the $3 \times 3$ submatrix in the right-bottom of $C$, the 
	transformation reads as:
	$$
	\spalignmat{
		1;
		h_1,1;
		0,h_1,1
	}
	\rightarrow 
	\spalignmat{
		1,*,0;
		h_1,0,0;
		0,h_1,1
	}
	\rightarrow
	\spalignmat{
		1,*,*;
		h_1,0,0;
		0,h_1,0
	}.
	$$
	
	\textbf{Case} \ref{item:case4}: In this case, we 
	first add $(-\frac{1}{h_1})$-times of the $(k-2)$-th column of $C$ to 
	the $(k-1)$-th column, and then swap the $(k-2)$-th column and 
	the $k$-th column. Focusing on the $3 \times 3$ submatrix in the 
	right-bottom 
	of $C$, the 
	transformation reads as:
	$$
	\spalignmat{
		1;
		h_1,1;
		h_2,h_1,1
	}
	\rightarrow 
	\spalignmat{
		1,*,0;
		h_1,0,0;
		h_2,0,1
	}
	\rightarrow
	\spalignmat{
		1,0,*;
		h_1,0,0;
		h_2,1,0
	}.
	$$
	
	\textbf{Case} \ref{item:case5}: In this case, we first 
	add the $(-\frac{1}{h_1})$-times of the $(k-2)$-th column to the 
	$(k-1)$-th column, and then add 
	the $(-\frac{1}{h_1-h_2/h_1})$-times of the $(k-1)$-th column to 
	the $k$-th column. The transformation reads as:
	$$
	\spalignmat{
		1;
		h_1,1;
		h_2,h_1,1
	}
	\rightarrow 
	\spalignmat{
		1,*,0;
		h_1,0,0;
		h_2,h_1-h_2/h_1,1
	}
	\rightarrow
	\spalignmat{
		1,*,*;
		h_1,0,0;
		h_2,h_1-h_2/h_1,0
	}.
	$$
\end{proof}
\begin{theorem}
Let $3 \le k \le n-2$, $\mathbf{a} = (a_1,\dots,a_n)$ be $n$ 
pairwise distinct evaluation points in $\F_q$ and 
$\mathbf{u} = (u_1,\dots,u_n)$. If $t_1 = 0$, then 
$\GRS_{k,2}(\mathbf{a},\infty)$ has a parity check matrix 
\begin{equation}
H_0 \defeq 
\spalignmat{
{G_{n-k+1,1}(\mathbf{a},\mathbf{u})}, \mathbf{c}_0^\intercal
}
\end{equation}
where $G_{n-k+1,1}(\mathbf{a},\mathbf{u})$ is the generator matrix of 
$\GRS_{n-k+1,1}(\mathbf{a},\mathbf{u})$ defined in 
\autoref{eqn:matrix_subcode_GRS}, 
and $\mathbf{c}_0 = (0,\dots,0,-1,t_2)$ is a vector of length $n-k+1$.
On the other hand, if $t_1 \neq 0$, then $\GRS_{k,2}(\mathbf{a},\infty)$ has 
a partiy check matrix 
\begin{equation}
H_1 \defeq 
\spalignmat{
{G_{+,n-k+1,\eta}(\mathbf{a},\mathbf{u})}, \mathbf{c}_1^\intercal
}
\end{equation}
where $G_{+,n-k+1,\eta}(\mathbf{a},\mathbf{u})$ is the generator matrix of 
the $(+)$-TGRS code $C_{+,n-k+1,\eta}(\mathbf{a},\mathbf{u})$ defined in 
\autoref{eqn:matrix_PTGRS}, 
$\eta = -1/t_1$ and $\mathbf{c}_1 = (0,\dots,0,-1,-t_2/t_1)$ is a 
vector of length $n-k+1$.
\end{theorem}
\begin{proof}
Let $G \defeq G_{k,2}(\mathbf{a},\infty)$ be the generator matrix 
of $\GRS_{k,2}(\mathbf{a},\infty)$ defined in 
\autoref{eqn:matrix_subcode_EGRS}. Let $\mathbf{g}_i, i \in J(k,2)$ be 
the row of degree $i$ in $G$, i.e., 
$\mathbf{g}_i = (a_1^i,\dots,a_n^i,0)$ for $i < k$ and 
$\mathbf{g}_k = (a_1^k,\dots,a_n^k,1)$. Similarly, let 
$\mathbf{h}_j, j \in J(n-k+1,1)$ be the rows of degree $j$ in $H_0$. 
It suffices to check that $\mathbf{g}_i \cdot \mathbf{g}_j = 0$ for all 
$i \in J(k,2)$, $j \in J(n-k+1,1)$. 

Ignoring the $0$ in the last entry of $\mathbf{g}_i$'s, we know that 
$\mathbf{g}_i, i \in J(k,2) \setminus \setnd{k}$ 
constitute a basis of $\GRS_{k-1,1}(\mathbf{a})$. As $4 \le k-1 \le n-4$ and 
$t_1 = 0$, by \autoref{theorem:dual_code_1}, the dual code of 
$\GRS_{k-1,1}(\mathbf{a})$ equals $\GRS_{n-k+1,1}(\mathbf{a},\mathbf{u})$. 
Therefore, 
$$
\mathbf{g}_i \cdot \mathbf{h}_j = 0, \, \forall i \in J(k,2)\setminus 
\setnd{k}, \forall j \in J(n-k+1,1).
$$
With \autoref{lemma:sum}, we can also check that 
$\mathbf{g}_k \cdot \mathbf{h}_j = 0$ for $j \in J(n-k+1,1)$. 
For the case $t_1 \neq 0$, the proof is similar as above, and detailes are 
omitted.
\end{proof}

\section{Conclusion}
In this paper, we considered a class of $[n,k]_q$ subcodes of the GRS codes, 
denoted as $\GRS_{k,r}(\mathbf{a},\mathbf{v})$. 
Equivalent characterizations for these codes being self-dual or NMDS were proposed. 
The self-dual property of $\GRS_{n/2,1}(\mathbf{a},\mathbf{v})$ is closely connected with 
the self-dual property of $\GRS_{n/2}(\mathbf{a},\mathbf{v})$, and known explicit constructions of 
self-dual $\GRS_{n/2}(\mathbf{a},\mathbf{v})$ can be exploited conveniently to the explicit construction of 
self-dual $\GRS_{n/2,1}(\mathbf{a},\mathbf{v})$, which is always self-dual NMDS or self-dual MDS. 
For $r \in \setnd{1,k-1}$ or $r = 2$, the dual codes were also found out. 
$\GRS_{k,1}(\mathbf{a},\mathbf{v})$ either is closed under taking dual codes, or results in $(+)$-TGRS codes by taking 
dual codes. In some cases, $\GRS_{k,2}(\mathbf{a},\mathbf{v})$ is also connected with the TGRS codes 
by taking dual codes. 

There are still some problems that might be interesting. For example, how to find 
more constructions of self-dual $\GRS_{n/2,n/2-1}(\mathbf{a},\mathbf{v})$, or can we 
exploit the known constructions for self-dual GRS codes to the construction of self-dual 
$\GRS_{n/2,n/2-1}(\mathbf{a},\mathbf{v})$ as in \autoref{corollary:self_dual_from_GRS}?
Also, for general $r$, what is the dual code of $\GRS_{k,r}(\mathbf{a},\mathbf{v})$ and are they 
connected with any known codes such as TGRS codes? 
It was shown that $\GRS_{k,r}(\mathbf{a},\mathbf{v})$ and $\GRS_{k,r}(\mathbf{a},\mathbf{v},\infty)$ 
can never be self-dual for $r \in [2,k-2]$. Can we modify these codes 
slightly so that they can be self-dual?
\label{sec:conclusion}

%\appendix
%\section{Proof of \autoref{theorem:self_dual_2}}
%\label{appsec:pf_self_dual_2}

\bibliographystyle{plain}
\bibliography{bib.bib}%bibtex entries
\end{document}